\documentclass[a4paper,11pt]{article}
\pdfoutput=1 % if your are submitting a pdflatex (i.e.~if you have
\usepackage{jheppub,amsthm} % for details on the use of the package, please
\usepackage{orcidlink}

\usepackage[utf8]{inputenc}
\usepackage{booktabs,multirow, quotes}
\usepackage[english]{babel}
\usepackage{amsmath,amssymb,graphicx,xcolor,alltt}
\usepackage{framed}
\usepackage{tcolorbox}
\usepackage{empheq}

\newcommand{\be}{\begin{equation}}
\newcommand{\ee}{\end{equation}}
\newcommand{\beq}{\begin{equation}}
\newcommand{\eeq}{\end{equation}}
\newcommand{\bea}{\begin{eqnarray}}
\newcommand{\eea}{\end{eqnarray}}

\renewcommand{\O}{\mathcal{O}}

\newcommand{\kI}{\kappa_{\rm I}}
\newcommand{\Ds}{\Delta}
\newcommand{\Depsilon}{\Delta_1}
\newcommand{\hD}{\hat \Delta_1}
\newcommand{\Imod}{I_{\rm model}}
\newcommand{\gmod}{g_{\rm model}}

\newcommand{\fk}{f_\textup{bulk}}
\newcommand{\fy}{f_\textup{bdy}}
\newcommand{\D}{\Delta}

\renewcommand{\Re}{\operatorname{Re}}
\renewcommand{\Im}{\operatorname{Im}}

\theoremstyle{remark}
\newtheorem{remark}{Remark}[section]

\theoremstyle{theorem}

\def\kI{\kappa_{\rm I}}

\makeatletter
\def\@fpheader{\ }
\makeatother

\title{Direct Experimental Test of Conformal Invariance via Grazing Scattering: A Proposal for X-ray and Neutron Experiments}

\author{Alessandro Podo\orcidlink{0000-0003-4166-3997},}
\author{Slava Rychkov\orcidlink{0000-0002-5847-1011}}

\affiliation{Institut des Hautes \'Etudes Scientifiques, 91440 Bures-sur-Yvette, France}

\emailAdd{podo@ihes.fr}
\emailAdd{slava@ihes.fr}

\abstract{We propose a test of conformal invariance in critical phenomena based on the study of a two-point correlation function in the presence of a boundary. This two-point function can be studied using X-ray or neutron scattering in the conditions of total reflection (so-called grazing scattering). The conformal Ward identity in momentum space is here expressed as a differential constraint on the scattering cross-section, as a function of the momentum transfer and the scattering angle. Experimental verification using X-rays and binary alloys appears well within the existing techniques, while feasibility for neutron scattering requires further study. This would be the first direct experimental test of conformal invariance in critical phenomena, a symmetry widely assumed but never directly verified. }

\begin{document} 
	
\maketitle
	
\section{Introduction}\label{sec:introduction}

Polyakov conjectured back in 1970 that fluctuations at continuous phase transitions possess conformal invariance~\cite{Polyakov:1970xd}. This was one of the motivations for the development of conformal field theory (CFT), both in 2D~\cite{Belavin:1984vu} and in higher dimensions~\cite{Polyakov:1974gs}. Nowadays, conformal invariance is routinely used to compute critical exponents~\cite{DiFrancesco:1997nk,Poland:2018epd}, in agreement with experimental measurements~\cite{Collins1989,Anisimov1991}. Quantum field theory understands conformal invariance as an emergent symmetry enhancing scale invariance, due to the generic tracelessness of the stress tensor at the renormalization group (RG) fixed point~\cite{Polchinski:1987dy,Nakayama:2013is}. Various predictions of conformal invariance were tested in numerical simulations~\cite{BCN,Langlands1994,Kennedy2018,Billo:2013jda,Cosme:2015cxa,Gori:2015rta,Zhu:2022gjc}. In some cases, conformal invariance of critical fluctuations has even been established as a mathematical theorem~\cite{Smirnov2006ICM}. From the theoretical point of view, there is no doubt that Polyakov's conjecture is correct. Surprisingly, to our knowledge, \emph{direct} experimental tests of conformal invariance are currently lacking. Conformal invariance in critical phenomena may be the only symmetry of nature which is firmly believed to be true theoretically, but not yet tested experimentally, at least not directly. Our goal here is to propose a realistic experiment which may fill this~gap.

We do mention experimental evidence of conformal random curve ensembles --- Schramm--Loewner evolution~\cite{Schramm2000} --- in planar turbulent motion of soap films \cite{Thalabard} and of living biological matter~\cite{Andersen2025}. Our proposal will concern instead a critical phenomenon with a theoretically well-established CFT description, staying close to the original Polyakov's conjecture.

Any agreement between a critical exponent computed by CFT methods (e.g.~the conformal bootstrap~\cite{Poland:2018epd}) and an experimental measurement of the same exponent constitutes an \emph{indirect} experimental test of conformal symmetry~\cite{Henkel1999,Rychkov:2025zks}. This may not be sufficiently appreciated. The exponents are also computed by several other methods, notably RG~\cite{Wilson:1973jj,Pelissetto:2000ek} or sometimes exact integrability~\cite{Baxter1982}. It may not always be clear what depends on what. In this respect we emphatically note that CFT is a self-sufficient framework, able to predict the exponents completely independently of any other technique. For many examples of agreements between CFT and experimental exponents see~\cite{Henkel1999,Rychkov:2025zks}. 

Such indirect tests notwithstanding, the test proposed below will be different because direct, in the sense that it will be testing the symmetry itself and not its consequences such as the values of the critical exponents. The direct meaning of symmetry is that the shape of an object remains the same after applying a symmetry transformation. In field theory, the objects on which the symmetry acts are correlation functions. Our direct test will be based on checking consequences of conformal symmetry for the functional form of correlation functions. 

The simplest such consequences were worked out by Polyakov back in 1970~\cite{Polyakov:1970xd}; they are well known. Two-point functions of so-called primary fields\footnote{These are fields which cannot be expressed as derivatives of other fields. See App.~\ref{sec:glossary} for a list of key theoretical concepts.} scale as a power with the distance and are nonzero only for operators of equal scaling dimensions. Three-point functions are determined up to a constant in terms of the scaling dimension of the fields. Four-point functions are determined up to a function of conformal cross-ratios. Experimentally, only two-point functions of lowest-dimension fields are easy to measure.\footnote{See however~\cite{Fang:2024uyf,Emperauger:2025raf,Sun:2026aqf} for recent studies of \emph{quantum} critical phenomena using Rydberg atoms where higher-point functions may be accessible in the future, and~\cite{RongRychkov} for a possible direct test conformal invariance in such experiments. Another observable which appears suitable for future tests of conformal invariance is the critical Casimir force~\cite{Hertlein2008,PhysRevE.80.061143}.} These are measured by scattering probes (X-rays or neutrons) off the critical sample. However, since the two-point function is a simple power, its functional form is fixed by scale invariance~alone.
 
How can we test conformal invariance, while remaining at the two-point function level? Pokrovsky~\cite{Pokrovskii1973,PatashinskiPokrovsky1979} proposed early on to test the vanishing of the two-point function between the two lowest primaries at the liquid-vapor critical point. To our knowledge the test was not carried out. Moreover, since this two-point function vanishes already due to the emergent $\mathbb{Z}_2$ invariance, this would not really test conformal symmetry.

\begin{figure}[h]
	\centering
	\includegraphics[width=0.9\textwidth]{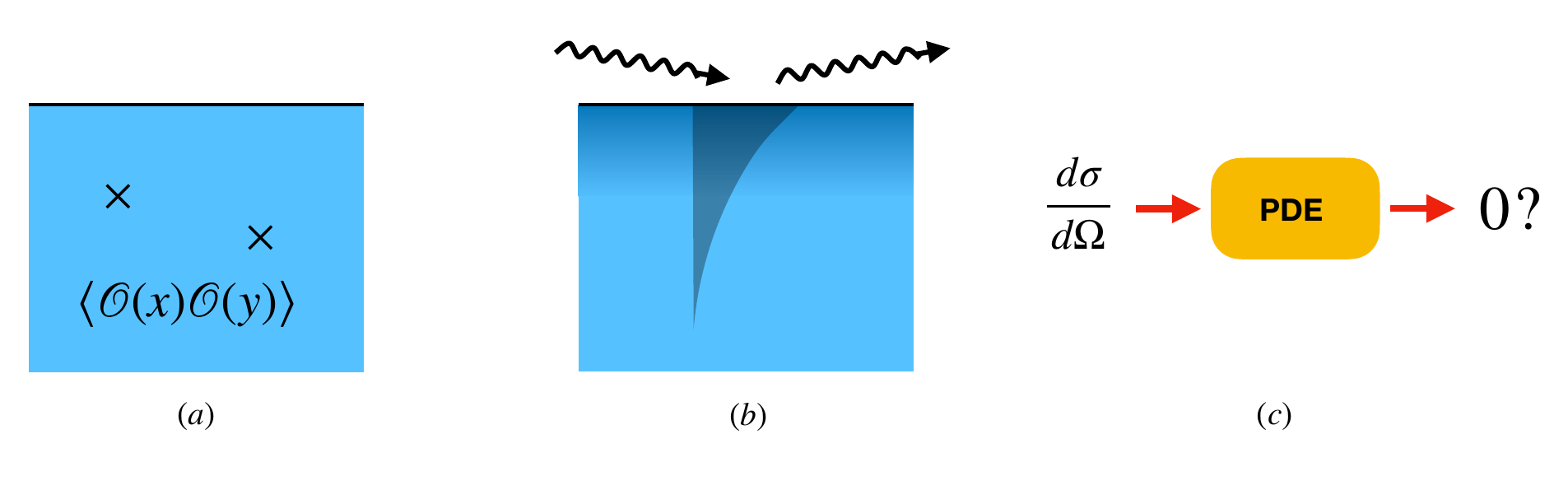}
	\vspace{-0.3cm}
	\caption{The main ideas of this paper: (a) Consider the two-point function of the fluctuating order parameter in presence of a flat boundary, whose near-boundary behavior is subject to kinematic constraints from conformal invariance. (b) Consider grazing scattering of an external probe (\mbox{X-ray}, neutron), which scatters on the fluctuations of the order parameter in the conditions of total reflection. The probe wavefunction decays exponentially into the sample, probing the near-boundary two-point function. (c) Write down the conformal invariance Ward identity in momentum space, as a partial differential equation on the cross-section as a function of momentum transfer and scattering angle. Applied to the scattering cross-section, this should give zero if conformal invariance is present.\label{fig:summary}}
\end{figure}

We will instead consider the two-point function of a single primary, but in a half-space, see Fig.~\ref{fig:summary}. It is well known that boundary critical phenomena have richer phenomenology than the pure bulk criticality~\cite{Diehl1986,Diehl:1996kd}. In particular, as noted by Cardy~\cite{Cardy:1984bb}, in this setting conformal invariance is more constraining than mere scale invariance. But how do we measure a boundary effect in the two-point function? Naively, the cross-section of a probe scattering will be dominated by the bulk. Here we will rely on the grazing-incidence small-angle scattering (GISAS)~\cite{WikiGISAS,PhysRevB.26.4146,PhysRevLett.51.1469,Dosch1992,Robinson1992,MullerBuschbaum2008}, or simply \emph{grazing scattering}. The probe is sent very nearly parallel to the surface of the sample, at an angle below the critical angle~$\alpha_c$, which depends on the sample's refraction index. Then, the probe cannot enter the bulk but experiences total reflection. The probe wavefunction decreases exponentially into the sample's bulk, and is sensitive to the two-point function of the fluctuations near the surface.  Our idea here will be to write the constraints imposed by conformal invariance directly in momentum space, as a differential operator acting on the cross-section of probe scattering. The latter is a function of the incidence and exit angles, and of the momentum transfer parallel to the boundary. This will facilitate comparison with experimental data, and is a crucial advantage with respect to a proposal of Gompper and Wagner~\cite{Gompper1985}, who first highlighted some consequences of conformal invariance on grazing scattering (see Remark~\ref{rem:Gompper}).

In Sections~\ref{sec:2pt_z} and~\ref{sec:grazing}, we will recall general results on two-point correlation functions in the presence of a plane boundary and describe how grazing scattering can be used to investigate boundary critical phenomena. We will then proceed to derive the consequences of conformal invariance on the probe scattering cross-section. Our main theoretical result is the conformal Ward identity in momentum space derived in Sec.~\ref{sec:conformal-ward-identity},  Eq.~\eqref{eq:C2g0}. Further consequences of this identity on the small momentum expansion of the critical scattering rate are studied in Sec.~\ref{sec:WIsmall}, before validating our results with a numerical study of the 3D ordinary Ising transition (Sec.~\ref{sec:num}). This allows us to present theoretical predictions on the observable grazing scattering rate and also check the conformal Ward identity in this numerical setting. 

In Section~\ref{sec:testing-conformal-invariance} we summarize the strategy of how conformal Ward can be tested in a realistic experimental setup. One needs to measure the scattering cross-section as a function of momentum transfer and scattering angles, then correct for Fresnel transmission coefficients, and finally apply the differential operator that we derived. This looks feasible with experimental techniques which exist since decades. We discuss X-ray grazing scattering experiments by Mail\"ander et al.~\cite{mailander1990near,mailander1990phase} from 1990. Their goal was to measure the scattering rate at small momentum transfer, confirming the ``cuspy'' behavior predicted by Dietrich and Wagner~\cite{PhysRevLett.51.1469}. They did not test our Ward identity which was not known at the time. Nor can we reuse their published data for such a test, because it was integrated over a range of outgoing scattering angles $\alpha_f$. However, repeating their experiment without integration in~$\alpha_f$ would yield a valid test. 

We conclude in Sec.~\ref{sec:conclusions} with an outlook and some open theoretical problems. Several technical results are presented in the appendices, among which a rigorous derivation of the small $p$ scaling laws for critical scattering at $T=T_c$ (App.~\ref{app:smallp}) and a detailed account of how the scattering rate can be obtained through numerical integration from known conformal data (App.~\ref{app:numint}). We also include a list of technical terms and abbreviations with brief explanations (App.~\ref{sec:glossary}).

\vspace{0.3cm}
\noindent {\bf Notation and conventions.} We consider Euclidean space, parametrized by coordinates ${x=(x_1,\ldots,x_{d})\in \mathbb{R}^{d}}$, and half-space, defined by $z=x_d\ge 0$, $y=(x_1,\ldots,x_{d-1})\in \mathbb{R}^{d-1}$. Greek indices $\mu,...$ refer to $\mathbb{R}^{d}$, while Latin indices $i,...$ refer to $\mathbb{R}^{d-1}$. In some theoretical considerations we keep $d$ general, while experimental tests refer to $d=3$.

\section{Two-point Function in Half-space}
\label{sec:2pt_z}

Our basic field-theoretic observable will be the critical two-point function of the order parameter (OP) in half-space. In later sections we will probe it with X-ray or neutron scattering. In this section we will recall how conformal invariance restricts the kinematics of this correlation function. We will also present what is known about it theoretically for the 3D ordinary Ising transition, which will be our primary example of experimental interest.

In the field-theoretic description of the critical point, we consider a scalar field $\O(x)$ having a scaling dimension $\Delta$, which is the continuum limit of the order parameter. In the jargon of CFT, ``fields'' are referred to as operators, and we will use these interchangeably. 

In full Euclidean space, the two-point function of $\O$ scales as a simple power:
\beq
\left\langle \O(x) \O(x')   \right \rangle_{\mathbb{R}^d}= \frac{1}{(x-x')^{2\Delta}}.
\label{eq:2pt}
\eeq
This two-point function describes how OP correlations decrease with the distance. For the 3D Ising transition, the operator $\O$ is the lowest $\mathbb{Z}_2$-odd field: its dimension $\Delta=0.518\ldots$ is known with $10^{-8}$ accuracy thanks to the conformal bootstrap; see Tab.~\ref{tab:3DIsing} below. The scaling dimension $\Delta$ determines the critical exponent $\eta$ by the well-known relation $\Delta=(d-2)/2+\eta/2$.

Eq.~\eqref{eq:2pt} is fixed by translation, rotation, and scale invariance, without the use of special conformal transformations. So we cannot use it to test for conformal invariance.

Consider instead the same correlation function in the half-space $z=x_d\ge 0$, $y=(x_1,\ldots,x_{d-1})\in \mathbb{R}^{d-1}$. Let us suppose that the theory in the bulk is conformally invariant. Let us also assume that the boundary conditions on the boundary $z=0$ are conformal, i.e.~they preserve the subgroup of conformal transformations leaving the boundary invariant. This is theoretically justified, because the RG flow will generically drive any microscopic boundary condition to a conformal boundary condition at long distances~\cite{Cardy:1984bb,Cardy1987}.\footnote{See~\cite{Nakayama:2012ed} for a discussion of subtle issues in this context.} Different choices of conformal boundary conditions correspond to different boundary universality classes. For example, ordinary, extraordinary and special 3D Ising universality classes correspond to different choices of conformal boundary conditions~\cite{Liendo:2012hy}.

It can be shown that the two-point function of a \emph{primary} (see below) conformal field in this geometry takes the form~\cite{Cardy:1984bb,McAvity:1995zd}
\beq
\left\langle \O(y,z) \O(y',z')  \right \rangle= \dfrac{1}{[(y -y')^2+(z-z')^2]^\Delta} G(\xi),
\label{eq:2ptbdry}
\eeq
where $G(\xi)$ is a function of the cross-ratio
\beq
\xi=\dfrac{(y -y')^2+(z-z')^2}{4z z'}\,,
\label{eq:xi}
\eeq
left invariant by the above-mentioned subgroup. The form of $G(\xi)$ cannot be fixed by conformal kinematics alone. We can say, though, that $G(\xi)\to 1$ as $\xi\to 0$, so that the bulk short-distance asymptotics of the two-point function are consistent with~\eqref{eq:2pt}. 

Let us say a few words about the requirement of being primary. Recall that fields in a conformal field theory can be primaries or descendants, which are derivatives of the primaries. The OP corresponds to the field of lowest scaling dimension in its symmetry class, which is therefore a primary, and the result~\eqref{eq:2ptbdry} applies.
 
Importantly, Eq.~\eqref{eq:2ptbdry} is more restricted than the form dictated by translation, rotation, and scale invariance, which without the use of conformal invariance would allow to consider two invariant cross-ratios:
 \beq
 \xi_1 = \dfrac{(y -y')^2}{4z z'},\quad \xi_2 =\dfrac{(z -z')^2}{4z z'}.
 \eeq 
The function $G$ could then depend on $\xi_1,\xi_2$ separately, and not only on their sum $\xi$ as in~\eqref{eq:2ptbdry}. 
Thus, with a plane boundary, already the two-point function can distinguish conformal invariance from the weaker scale+translation+rotation symmetry~\cite{Cardy:1984bb}.

\subsection{$G(\xi)$ for 3D Ordinary Ising Transition}\label{sec:gxi-for-3d-ordinary-ising-transition}

Although conformal kinematics alone cannot fix $G(\xi)$, CFT has other conditions, most importantly that the boundary and bulk operator product expansions (OPE) have to be consistent. Once this is imposed, $G(\xi)$ can be determined, through a procedure known as the conformal bootstrap. In this section we will review this and describe $G(\xi)$ for the 3D ordinary Ising transition. This will be needed in particular in the numerical test of the conformal Ward identity, Section~\ref{sec:num}.

\begin{figure}[h]  
	\centering
	\includegraphics[width=0.6\textwidth]{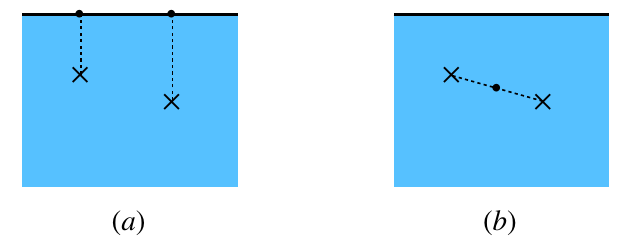}
	\caption{Schematic representation of (a) boundary OPE and (b) bulk OPE.}
	\label{fig:ope}
\end{figure}

Our setup is still that of a field theory in half-space. The boundary OPE expansion, Fig.~\ref{fig:ope}(a), replaces every operator $\O(x)$ inside any correlation function by an expansion of the form:\footnote{The factor of 2 in $2z$ is conventional, to match the conventional factor of 4 in the denominator of \eqref{eq:xi}.}
\beq
\O(y,z) = \sum_{i=1}^\infty \mu_i (2z)^{\hat{\Delta}_i-\Delta} \hat{\O}_i(y) \, .
\label{eq:b2b0}
\eeq
Here $\hat{\O}_i$ are operators which live on the boundary $z=0$. They are characterized by their own scaling dimensions $\hat{\Delta}_i$. The $z$ behavior of the prefactors in~\eqref{eq:b2b0} respects scale invariance. The numerical ``OPE coefficients'' $\mu_i$ depend on the theory but not on the correlation function. Using this OPE inside the two-point correlation function of $\O$, we can represent it as a double infinite sum
\beq
\langle\O(y,z)  \O(y',z')\rangle =  \sum_{i,i'=1}^\infty \mu_i \mu_{i'} (2z)^{\hat{\Delta}_i-\Delta} (2z')^{\hat{\Delta}_i-\Delta}\langle \hat{\O}_i(y) \hat{\O}_{i'}(y')\rangle\,.
\label{eq:b2bexp}
\eeq
This holds in any theory, conformal or just scale invariant. But in a CFT more is true. We can group the boundary operators into conformal multiplets: the primary boundary operators $\hat{\O}_i$ and their derivatives along the boundary. As a consequence of conformal invariance, there is one independent coefficient $\mu_i$ per multiplet, namely that of the primary. Furthermore, fields in two different multiplets have zero two-point functions. Resumming the expansion~\eqref{eq:b2bexp}, the two-point function agrees with~\eqref{eq:2ptbdry} and, moreover, one gets a formula representing $G(\xi)$ as an expansion in ``boundary conformal blocks'' $\fy$~\cite{McAvity:1995zd}:\footnote{For the case of interest $d=3$ the boundary blocks are elementary functions:
	$\fy(\hat\D,\xi)\vert_{d=3}=\frac1{2\sqrt{\xi}}
	[4/({1+\xi})]^{\D-\frac12}[1+
	\sqrt{\xi/(1+\xi)}]^{-2(\D-1)}
	$.}\footnote{Sometimes one has to add the term $\xi^{\D}\mu_0^2$ in the r.h.s., corresponding to the unit boundary operator. This term is absent for the ordinary transition.}
\begin{align}
G(\xi) &= \xi^{\D} \sum_{i=1}^\infty \mu_{i}^2 \,\fy(\hat{\D}_i,\xi) \,,
\label{eq:crossingsy0}\\
\fy(\hat\D,\xi) &= \xi^{-\hat\D} \,
{}_2F_1\left(\hat\D,\hat\D+1-\frac{d}{2};2\hat\D+2-d;-\frac{1}{\xi}\right).
\label{eq:crossingsy0-block}
\end{align}

Similarly, the bulk OPE corresponds to the limit when two operators approach each other instead of the boundary,  Fig.~\ref{fig:ope}(b). In the r.h.s. one finds a series of bulk operators $\O_k$ times OPE coefficients $\lambda_k$. This reduces the two-point function of $\O$ to an infinite sum of one-point functions of $\O_k(y,z)$ which by scale+translation invariance have the form
\beq
\langle \O_k(y,z)\rangle = a_k (2z)^{-\Delta_k}\, .
\eeq
The ``one-point function coefficients'' $a_k$ are another set of constants characterizing the theory. In a CFT, this computation can be pushed further grouping bulk fields in conformal multiplets and using constraints of conformal invariance on relative factors among OPE coefficients and one-point function coefficients. One then obtains a second expansion of~$G(\xi)$, this time in ``bulk conformal blocks'' $\fk$~\cite{McAvity:1995zd}:
\begin{align}
	G(\xi) &=  1+ \sum_{k=1}^\infty \lambda_{k} a_k \,\fk(\D_k,\xi)  \,,
	\label{eq:crossingsy1}\\
	\fk(\D,\xi) &= \xi^{\D/2} \,
	{}_2 F_1 \left( \frac{1}{2}\D, \frac{1}{2}\D; 
	\D +1-\frac{d}{2};-\xi \right) .
\end{align}

Although the expansion~\eqref{eq:crossingsy0} converges most rapidly for $\xi\gg 1$ while~\eqref{eq:crossingsy1} does so for $\xi\ll 1$, both of these expansions converge for all $0<\xi<\infty$. Imposing their equality in this range, and in particular around $\xi\sim 1$, gives a nontrivial constraint on the parameters characterizing the CFT ($\Delta_i$, $\hat{\Delta}_i$, $\mu_i^2$, $\lambda_k a_k$, etc.), and on the function $G(\xi)$ itself. Investigating such constraints is what the conformal bootstrap is about.

In Tab.~\ref{tab:3DIsing} we report the values of several parameters yielded by conformal bootstrap studies of the ordinary 3D Ising transition:
\begin{itemize}
	\item
The OP dimension $\Delta$;
\item 
The dimensions $\Delta_k$ and the products $\lambda_k a_k$ for the first two bulk OPE scalar primaries;
\item
The dimension $\hat{\Delta}_1$ and coefficient $\mu_1$ of the leading primary in the boundary OPE.
\end{itemize}

\begin{figure}[h]  
	\centering
	\includegraphics[width=0.6\textwidth]{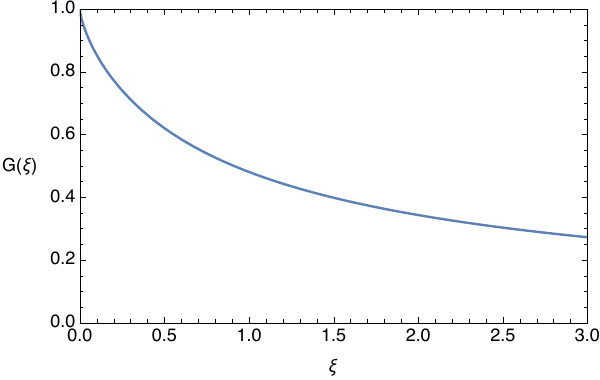}
	\caption{The function $G(\xi)$ for the ordinary boundary transition of the 3D Ising model.}
	\label{fig:GIsing}
\end{figure}

Using this table, we compute $G(\xi)$ numerically as follows. For $\xi\le 1$ we use the bulk expansion~\eqref{eq:crossingsy1} truncated to two nontrivial terms $k=1,2$. For $\xi\ge 1$ we use the boundary expansion~\eqref{eq:crossingsy0} truncated to a single term $i=1$. We use central values for all parameters except for $\mu_1^2$, which we adjust to have continuous gluing at $\xi=1$. This requires $\mu_1^2=0.753888$, well within the error in the table. Perhaps surprisingly, this simple procedure already gives an excellent approximation for all $0<\xi<\infty$. 
The check is that the two expansions agree very closely at $\xi\sim1$ and not just at $\xi=1$. The resulting function~$G(\xi)$, Fig.~\ref{fig:GIsing}, is continuous for $0<\xi<\infty$. This approximation to $G(\xi)$ will be used in the $R(u,y)$ integral in Section~\ref{sec:num} and App.~\ref{app:numint}. Note in particular the asymptotics
\beq
G(\xi) =
\begin{cases} 1 - \lambda_1 a_1 \xi^{\Depsilon/2}+\ldots,& \xi \to 0,\\
	\mu_1^2 \xi^{\Ds -\hat \Delta_1}+\ldots, & \xi \to \infty\,.
\end{cases}
\eeq

\begin{table}
	\centering
\begin{tabular}{lll}
	\toprule
	Order Parameter & $\Ds=0.518148806(24)$~\cite{Chang:2024whx} & \\
	\midrule
	Bulk OPE & $\Depsilon = 1.41262528(29)$~\cite{Chang:2024whx} & $\lambda_1 a_1 = -0.789(3) $~\cite{Gliozzi:2015qsa}\\
	& $\Delta_2 = 3.82951(61)$~\cite{Reehorst:2021hmp} & $\lambda_2 a_2 =0.042(1) $~\cite{Gliozzi:2015qsa}\\
	\midrule
	Boundary OPE & $\hat \Delta_1= 1.276(2)$~\cite{Gliozzi:2015qsa} & $\mu_1^2=0.755(13)$~\cite{Gliozzi:2015qsa}\\
	\bottomrule
\end{tabular}
\caption{A subset of parameters characterizing the 3D ordinary Ising transition, important for the evaluation of $G(\xi)$. The CFT describing this transition has not been exactly solved. We quote the current best estimates from the conformal bootstrap. For Ref.~\cite{Gliozzi:2015qsa}, see their Table 1 ($N=1$). Although they used older estimates of bulk CFT parameters as an input, this is likely a subleading source of error. In our calculation of $G(\xi)$ we used $\mu_1^2=0.753888$ to get a continuous gluing at $\xi=1$, while using central values for all other parameters.}
\label{tab:3DIsing}
\end{table}

\section{Grazing Scattering}
\label{sec:grazing}

An important experimental technique to investigate critical phenomena is that of scattering probes off the sample, which may be done with X-rays or neutrons~\cite{Dosch1992,dietrich1995scattering}. One uses neutron scattering to study ferromagnetic phase transitions, and X-rays scattering to study non-magnetic phase transitions, such as in binary alloys. 

In this section we will discuss scattering precisely at the critical point, i.e.~at $T=T_c$. (For completeness, scaling laws close to the critical point are reviewed in App.~\ref{app:scaling}.) After reviewing well-known predictions for the critical scattering rate in an infinite sample, we will discuss grazing scattering in half-space, which is our main interest. The goal is to recall the relation expressing the scattering rate in terms of the field theory two-point function in half-space, Eq.~\eqref{eq:grazing-sigma}. The main object of further study is the ``theory scattering rate'' defined by stripping the Fresnel factor from this relation. We describe the salient features of the theory scattering rate in the small $p$ expansion, including the characteristic ``cuspy'' behavior which was tested experimentally in 1990. 

The material in this section is mostly well known \cite{Dietrich1984,Dosch1992,Robinson1992}, but necessary to prepare the ground for the conformal Ward identities. We will focus on X-rays and binary alloys, but a similar discussion applies to neutron scattering as well, see Remark \ref{rem:neutron}.

\subsection{Infinite Sample}\label{sec:infinite-sample}
Recall first what happens in infinite sample without taking boundary effects into account. A probe propagating in the sample scatters on a nonuniform potential $V(x)$, where $V$ is some microscopic fluctuating quantity (magnetic moment or electron density), drawn from an equilibrium statistical distribution. This potential is considered fixed over the duration of time that the probe spends inside the sample. However the next scattered probe feels a different instance of $V(x)$. In the Born approximation, the cross-section for initial 3-momentum $K_i$ and final momentum $K_f$ is proportional to the squared matrix element of $V(x)$ between these two plane waves,
\beq
\frac{d\sigma}{d\Omega}\propto \left\langle\left|\int d^3x \psi_f^*(x) V(x) \psi_i(x) \right|^2\right\rangle\, ,
\label{eq:scat-rate}
\eeq
where $\psi_i(x)=e^{iK_ix}$, $\psi_f(x)=e^{iK_fx}$. Here $\langle\ldots\rangle$ denotes the statistical average over $V(x)$. Using translation invariance, this can be equivalently written as
\beq
\frac{d\sigma}{d\Omega}\propto \int d^3x e^{iQ x}  \langle V(x) V(0) \rangle \, ,
\eeq
where $Q =K_i-K_f$ is the momentum transfer. Note that $|K_i|=|K_f|=K$ as we consider only elastic scattering.

At the critical point, the long distance behavior of the microscopic correlation function $\langle V(x) V(0) \rangle$ is, up to a constant, well represented by the CFT correlator $\langle \mathcal{O}(x) \mathcal{O}(0) \rangle$ where $\mathcal{O}(x)$ is the CFT operator of lowest scaling dimension with which $V(x)$ has nonzero overlap. Using Eq.~\eqref{eq:2pt}, we predict the scaling of the scattering cross-section at low $Q$:
\beq
\frac{d\sigma}{d\Omega}\propto 1/|Q|^{3-2\Delta}\,.
\label{eq:as-sigma}
\eeq
One remark is in order. In the above discussion we assumed a ferromagnetic transition so that $V(x)=C \mathcal{O}(x)+\ldots$. At an antiferromagnetic-type transition we will have instead 
\beq
V(x)=C e^{-i Q_0 x}\mathcal{O}(x)+\ldots,
\label{eq:match}
\eeq 
where $Q_0$ is a nontrivial momentum in the Brillouin zone (``superlattice Bragg momentum''). To cancel the oscillating factor, we then consider probes scattering with momentum transfer $Q=Q_0+P$ with $P$ small, and the asymptotic behavior~\eqref{eq:as-sigma} holds with $Q$ replaced by $P$. 

\subsection{Half-space}\label{sec:half-space}

\begin{figure}
\centering
\includegraphics[width=0.9\textwidth]{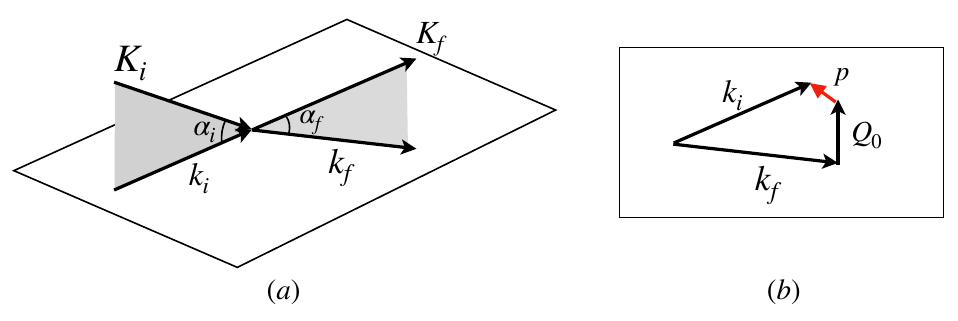}
\caption{Kinematics of surface scattering (adapted from~\cite{Dietrich1984}). (a) Incoming and outgoing X-ray momenta $K_i$ and $K_f$ and their projections on the surface $k_i$, $k_f$. (b) Projection on the surface. $Q_0$~is the superlattice Bragg momentum. We assume $Q_0$ to be parallel to the boundary, Eq.~\eqref{eq:Q0}.\label{fig:kinematics}}
\end{figure}

Let us now examine scattering off a half-space sample located at $z>0$. The probe arrives from/departs into the empty half-space $z<0$ with 3-momenta $K_i$ and  $K_f$. We denote by $k_{i}$,$k_f$ the 2-dimensional projections of $K_{i},K_f$ on the boundary, and by $\alpha_{i},\alpha_f$ the incidence and exit angles (see Fig.~\ref{fig:kinematics}). Because of refractive phenomena on the boundary, the incoming and outgoing probe wavefunctions inside the sample are given by ($x=(y,z)$)
\beq
\psi_i(x)= e^{i(k_i y+\kappa_i z)},\quad \psi_f(x)=e^{i(k_f y+\kappa_f z)} \, .
\eeq
We describe the refraction in the homogeneous medium approximation via the refractive index $n$, assumed $n<1$. Introducing the critical angle $\alpha_c$, $\cos\alpha_c =n$, we have
\beq
\kappa_i = K(\sin^2\alpha_i -\sin^2 \alpha_c)^{1/2}\,, \quad \kappa_f= -K(\sin^2\alpha_f -\sin^2 \alpha_c)^{1/2}\,.
\eeq
Note that $\kappa_i, \kappa_f$ are real for incidence and exit angles $\alpha_i,\alpha_f >\alpha_c$ but become imaginary for $\alpha_i,\alpha_f<\alpha_c$. The sign of the imaginary part is then chosen so that the wavefunction exponentially decays inside the sample.
 
In terms of these wavefunctions, the scattering cross-section is given by the same formula~\eqref{eq:scat-rate} as before, restricting integration to $z>0$. Using translation invariance along the boundary, we can write this as~\cite[(2.14)]{Dietrich1984}
\beq
\label{eq:dsigma}
\frac{d\sigma}{d\Omega}= \mathcal{F} \int_{z,z'>0} d^2y\, dz\, dz'\, e^{iq y+i (\kappa z -\kappa^* z')}  \langle V(y,z) V(0,z') \rangle \, ,
\eeq
where $q =k_i-k_f$ is the momentum transfer along the boundary and 
\beq
\label{eq:kappa}
\kappa=K(\sin^2\alpha_i -\sin^2 \alpha_c)^{1/2}+ K(\sin^2\alpha_f -\sin^2 \alpha_c)^{1/2}\,.
\eeq
If both $\alpha_i,\alpha_f >\alpha_c$ then $\kappa$ is real and the cross-section integral is dominated by the bulk. Our interest is instead in the grazing scattering case, which arises when one or both of $\alpha_i,\alpha_f$ are subcritical, so that the imaginary part of $\kappa$ is positive: $\Im \kappa>0$. The scattering integral is then localized near the boundary. 

The ``Fresnel factor'' $\mathcal{F}$ in~\eqref{eq:dsigma} involves, among other things, transmission coefficients given by the Fresnel formulas in terms of $\alpha_i,\alpha_f$~\cite[App.~A]{Dietrich1984}.\footnote{In binary alloys, the proportionality factor $\mathcal{F}$ also carries a dependence from the (difference of the) atomic form factors of atoms A and B~\cite{Dietrich1984}. This is expected to be a small effect unless the X-ray frequency is close to that of an absorption or emission line. If needed, this dependence can be corrected by using the experimentally measured atomic form factors. \label{atomicFF}}

Using the proportionality rule~\eqref{eq:match} with a constant $C$, we now replace the microscopic two-point function by the field theory two-point function $\langle\mathcal{O}(y,z) \mathcal O(0,z')\rangle$ discussed in Sec.~\ref{sec:2pt_z}. Importantly, if the phase transition is antiferromagnetic we will assume that 
\beq 
\text{the vector $Q_0$ in~\eqref{eq:match} is parallel to the boundary $z=0$.}\label{eq:Q0}
\eeq 
(This requires cutting the sample in a particular way.) Then we obtain the final result
\begin{align}
\frac{d\sigma}{d\Omega}&= \mathcal{F} C^2 \, \mathcal{G}(p,\kappa)\,,\nonumber\\
\mathcal{G}(p,\kappa) &= \int_{z,z'>0} d^2y\, dz\, dz'\, e^{i p y+i (\kappa z -\kappa^* z')}  \langle \mathcal O(y,z) \mathcal O(0,z') \rangle\,,
\label{eq:grazing-sigma}
\end{align}
corresponding to $q= Q_0+p$ in Eq.~\eqref{eq:dsigma}. Experiments probing critical behavior are performed at $p$ small. 

The ``theory scattering rate'' $\mathcal{G}(p,\kappa)$ will be our main object of interest. Experimentally, we measure the true scattering rate $\frac{d\sigma}{d\Omega}$ and divide by the Fresnel factor $\mathcal{F}$, whose dependence on the kinematic variables is known as mentioned. This  gives $\mathcal{G}(p,\kappa)$ up to the normalization factor $C^2$. This factor cannot be easily fixed, but it is a constant independent of the angles or $ p$. We will only make statements which do not depend on this normalization factor.

Since $\Im\kappa>0$, let us parametrize $\kappa = i|\kappa| e^{i\phi}$ with a real phase $\phi$, $|\phi|<\pi/2$. Using scale invariance and rotation invariance in $p$, we write the theoretical scattering rate as
\beq
\mathcal{G}(p,\kappa) = |\kappa|^{2\Delta-4} g(|p|/|\kappa|, \phi)\,.
\label{eq:Gscale}
\eeq
In other words, we use $|\kappa|$ as the typical momentum scale.\footnote{One could also use $\Im\kappa$ for this, but this would obscure the simple $\phi$ dependence of non-analytic terms.} 

The two-point function in the presence of a boundary does not typically have a simple analytic form. As discussed in the previous section, in a CFT it can be parameterized in terms of a single function of a cross-ratio $G(\xi)$, Eq.~\eqref{eq:2ptbdry}. Sometimes, like for the 3D ordinary Ising transition, this function $G(\xi)$ is known with good accuracy thanks to the conformal bootstrap (see Sec.~\ref{sec:gxi-for-3d-ordinary-ising-transition}). The integral~\eqref{eq:grazing-sigma} can then be done numerically, see Sec.~\ref{sec:num}. 

Furthermore, it is possible to perform a small $p$ expansion, which takes the following form (see App.~\ref{app:smallp}) 
\beq
g(p,\phi)= \sum_{i,j=1}^N a_{ij}(\phi) p^{\hat\Delta_i+\hat\Delta_j-2} + \sum_{k=0}^{k_N-1} e_k(\phi) p^{2k}+O(p^{2k_N}) \, .
\label{eq:gpphi}
\eeq
The main idea is to use the boundary OPE expansion~\eqref{eq:b2bexp} and integrate term by term. However the resulting expansion is only asymptotic. So in a more careful treatment (see App.~\ref{app:smallp}) we replace the exact two-point function by the sum of the first $N$ terms of the boundary OPE expansion, plus the error term. The integral of the boundary OPE terms can then be done explicitly, it gives the first sum in~\eqref{eq:gpphi} consisting of terms which behave as non-integer powers of $p$; we call them non-analytic. On the other hand the error term integral gives the second sum in~\eqref{eq:gpphi} consisting of terms which behave at small $p$ as a sum of integer powers of $p^2$. We call these terms analytic, although we do not claim that the series converges as $k_N\to \infty$. The expansion is asymptotic with an error $O(p^{2k_N}) $ where $k_N$ depends on the number of terms in the boundary OPE that we subtracted at the beginning, namely $k_N=\lfloor\hat\Delta_{N+1}\rfloor-1$. The coefficients in the expansion~\eqref{eq:gpphi} are functions of $\phi$. The coefficients $a_{ij}(\phi)$ have an interesting property (see App.~\ref{app:smallp}) $a_{ij}(\phi) \propto e^{i\phi(\hat\Delta_j-\hat\Delta_i) }$.
In particular diagonal coefficients $a_{ii}$ are constants. We are not aware of any simple properties of the coefficients $e_k(\phi)$.

The expansion~\eqref{eq:gpphi} applies to a general scale invariant theory, as long as the boundary OPE holds. Once we develop our conformal Ward identity, we will see that in a CFT there are additional relations which constrain the dependence on $\phi$ of the coefficients of different terms in the expansion.

\begin{figure}
\centering
\includegraphics[width=0.4\textwidth]{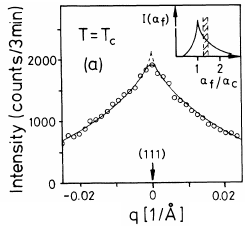}
\caption{Experimental measurement of the X-ray scattering intensity as a function of $|p|$ (denoted $q$ in the figure) in Fe$_3$Al binary alloy, reproduced from Refs.~\cite{mailander1990near,mailander1990phase} and showing the behavior~\eqref{eq:asymp0phi}. See Section~\ref{sec:testing-conformal-invariance} for further discussion of these data. \label{fig:Mailander}}
\end{figure}

In the simplest case of~\eqref{eq:gpphi}, we retain the first boundary operator $N=1$ and the zeroth analytic term:
\begin{align}
\label{eq:asymp0phi}
g(p,\phi) &= e_0(\phi) + a_{11}{p}^{2\hat\Delta_1-2} +\ldots,\\
&= e_0(\phi) + a_{11}{p}^{0.552} +\ldots,
\end{align}
where $e_{0}(\phi)$ is a nontrivial function (see Sec.~\ref{sec:num}), $a_{11}$ is a constant independent of $\phi$, and in the second line we used $\hat\Delta_1$ from Tab.~\ref{tab:3DIsing} for the ordinary 3D Ising transition. In the latter case $a_{11}$ is also negative, see App.~\ref{app:smallp}. Thus, equation~\eqref{eq:asymp0phi} predicts a characteristic ``cuspy'' behavior of the scattering rate at small $p$~\cite{PhysRevLett.51.1469}. This behavior was experimentally verified in Ref.~\cite{mailander1990near,mailander1990phase,burandt1993near}, see Fig.~\ref{fig:Mailander} and Sec.~\ref{sec:testing-conformal-invariance}.

\begin{remark}\label{rem:neutron} The analysis of elastic neutron scattering proceeds along similar lines~\cite{Dietrich1985,Dosch1992}. Magnetic scattering from antiferromagnets appears particularly suitable for the study of the cuspy behavior of the elastic cross section. In this case the factor $\mathcal{F}$ depends on transmission coefficients whose explicit dependence from the scattering angles can be found in Ref.~\cite{Dietrich1985}, and possibly on the magnetic form factors, whose momentum dependence is expected to be negligible away from transmission or emission lines. We note that conformal invariance is a property at equilibrium (statical critical behavior). Therefore we do not discuss inelastic magnetic scattering, which is used to probe the dynamic critical behavior. 

\end{remark}

\section{Theory: Conformal Ward Identity}\label{sec:conformal-ward-identity}

This section derives the main new theoretical result of the paper: the momentum-space conformal Ward identity. It is a partial differential operator which, acting on the theory scattering rate, should give zero if the theory is conformally invariant. Recall that the theory scattering rate is an integral transform of the two-point function in half-space, which can be obtained by dividing the experimentally accessible scattering rate by the Fresnel factor.

To check conformal symmetry, we have in principle two choices. The first strategy would be to perform detailed measurements of the scattering rate as a function of all parameters, invert the integral transform, reconstruct the position-space two-point function, and see if it has the form~\eqref{eq:2ptbdry}. That is clearly very laborious. Imposing the constraint directly on the theory scattering rate, i.e.~in momentum space, is clearly more efficient. 

As is well known, constraints in position space can be equivalently expressed in functional form such as~\eqref{eq:2ptbdry}, or in infinitesimal (differential) form. In field theory terminology the latter are referred to as Ward identities (WIs). The position space conformal WI is well known. Applying to it the integral transform, we will find our momentum-space conformal~WI.  

The actual derivation is structured as follows. In Section~\ref{sec:full-space-reminder} we recall how momentum-space WIs follow from position-space WIs in full space. Then, we will derive the momentum-space WI in half-space. It is convenient to split it into two steps. First, in Section~\ref{sec:half-space} we will do this for the auxiliary function $G(p,\kappa ,\kappa') $ obtained replacing $\kappa^*$ in the definition of $\mathcal G(p,\kappa)$ by an independent variable $\kappa'$. Then, in Section~\ref{sec:restriction-to-the-submanifold-kappakappa} we show that the WI restricts nicely to the submanifold $\kappa'=\kappa^*$ and find this restriction, Eq.~\eqref{eq:C20}. This WI, and Eq.~\eqref{eq:C2g0} re-expressing it in terms of the function $g(p,\phi)$, are the main results.

\subsection{Reminder: Ward Identity in Full Space}\label{sec:full-space-reminder}

CFT correlators are usually studied in position spaces. However, sometimes it is interesting or useful (like in the present context) to transform them to momentum space, e.g.~\cite{Polyakov:1974gs,bzowski2014implications,Coriano:2013jba,Gillioz:2018mto,Gillioz:2019lgs,Bzowski:2019kwd,Gillioz:2019iye,Gillioz:2020mdd,Gillioz:2020wgw}. 
Here we are interested especially in how constraints of conformal invariance manifest itself in momentum space. 
Let us first recall how these ``momentum space WIs'' would be derived for the case of CFTs without boundaries~\cite{Maldacena:2011nz,bzowski2014implications,Gillioz:2022yze}. 
Position-space WIs expressing the conformal invariance of the two-point function $\langle\O(x_1)  \O(x_2) \rangle$ take the form\footnote{See~\cite{DiFrancesco:1997nk,Rychkov:2016iqz,Simmons-Duffin:2016gjk} for pedagogical explanations.}
\be
\langle [ L, \O(x_1) ] \O(x_2) \rangle + \langle \O(x_1) [ L, \O(x_2)] \rangle =0\, ,
\label{eq:posWI}
\ee
where $L$ is any of the generators of the conformal group, and $[ L, \O(x) ]$ is a commutator, given by a differential operator acting on $\O(x)$ (see below). Conformal group generators include translation, rotation, scale transformation, as well as special conformal transformation (SCT) generators. The latter distinguish the conformal group from its smaller subgroup consisting of translations, rotations, and scale transformations alone. We will then consider the scale generator $D$ and the SCT generator $K_\mu$. They act on the CFT scalar primary operator $\O(x)$ of dimension $\Delta$ as (we use the sign conventions of~\cite{Simmons-Duffin:2016gjk})
\begin{align}
	[ D, \O(x)] &= \left(\Delta + x\cdot \partial \right) \O(x)  \, , \\
[ K_\mu, \O(x)] &= \left(2\Delta \;x_\mu   + 2 x_\mu (x\cdot \partial) - x^2 \partial_\mu \right) \O(x)  \, .
\end{align}
The corresponding position-space WIs~\eqref{eq:posWI} are written explicitly as:
\begin{align}
\text{Scale: } &\sum_{x=x_1,x_2} \left(\Delta+ x_\nu \partial_{x_\nu}\right)\langle \O(x_1) \O(x_2) \rangle  =0 \label{eq:scale-x} \, ,\\
\text{Conf: }&\sum_{x=x_1,x_2} \left(2 \Delta\; x_{\mu}+ 2 x_{\mu} x_\nu \partial_{x_\nu}- x^2 \partial_{x_\mu}\right)\langle \O(x_1) \O(x_2) \rangle  =0 \, .\label{eq:conf-x}
\end{align}
Let us now transform these WIs to momentum space. Because of translation invariance, we can set $x_2=0$. We then Fourier-transform with respect to~$x_1=x$, and express the result as a differential operator acting on the Fourier transform of the two-point function:
\beq
G(P) = \int d^d x e^{i P x}\langle \O(x) \O(0) \rangle\,.
\eeq
As we have seen in Section~\ref{sec:infinite-sample}, the probe scattering rate is proportional to this Fourier transform.

The resulting momentum-space WIs have the form~\cite{bzowski2014implications}:\footnote{
	Under our convention for Fourier transform we have $x_\mu f(x) \to -i \partial_{P_\mu} f(P)$ and $\partial_{x_\mu} f(x) \to -i P_\mu f(P)$. Thus, the l.h.s.~of the conformal WI with $x_1=x$, $x_2=0$, maps by Fourier transform to $-i$ times (notice the order of operations) $
	\left(2 \Delta \partial_{P_\mu}  -  2\, \partial_{P_\mu} \partial_{P_\nu}  P_\nu + 
	 (\partial_{P_\nu})^2 P_\mu \right)G(P) 
	$. Using the commutator $[\partial_{P_\nu},P_\mu]=\delta_{\nu\mu}$ a few times we get~\eqref{eq:momWI}. The scale WI is similar.} 
\begin{align}
	\text{Scale: } &\left[2\Delta-d  -  \,P_\nu \partial_{P_\nu}  \right] G(P) =0\,,\\
	\text{Conf: } &\left[2 (\Delta-d) \partial_{P_\mu}  -  2 \,P_\nu \partial_{P_\nu} \partial_{P_\mu} + P_\mu \partial_{P_\nu} \partial_{P_\nu} \right] G(P) =0 \, .
\label{eq:momWI}
\end{align}
For the case of a two-point function of a scalar primary operator in full space, we have $G(P)=G(|P|)$, by rotation invariance. Then the scale WI can be written as:
\beq
(2\Delta-d  -  |P| \partial_{|P|} )G =0\,.
\eeq
The conformal WI is $\partial_{|P|}$ of this. As noted in Section~\ref{sec:2pt_z}, in this case conformal invariance does not add anything new. According to the discussion there, we do expect to find additional constraints in half-space. Let us see how this happens.

\subsection{Ward Identity in Half-space I}\label{sec:half-space}

We have to study the integral in the r.h.s.~of~\eqref{eq:grazing-sigma}. In this section we will temporarily relax the constraint that $\kappa^*$ is the complex conjugate of $\kappa$. We thus consider an auxiliary function $G(p,\kappa ,\kappa')$ given by the integral (in the general $d$-dimensional case)
\be 
\label{correlator_bdy}
G(p,\kappa ,\kappa') = \int_0^\infty dz \, \int_0^\infty dz' \, e^{i (\kappa z - \kappa' z')} \int d^{d-1}y \,e^{i p y}  \left\langle \O(y,z) \O(0,z') \right\rangle  ,
\ee
as a function of $p$ and of two complex numbers $\kappa,\kappa'$, assuming $\Im \kappa>0$, $\Im \kappa'<0$. Since $\kappa,\kappa'$ are independent, we will be able to differentiate in them independently. Later in Section~\ref{sec:restriction-to-the-submanifold-kappakappa} we will have to restrict to the submanifold $\kappa'=\kappa^*$.

We are not aware of many prior works considering boundary CFT correlators in momentum space. Refs.~\cite{Prochazka:2018bpb,Prochazka:2019bhv} are one example, but with a very different setting and motivations.

Let us discuss the scale WI first. Separating coordinates into $y$ and $z$, and setting $y_2=0$, Eq.~\eqref{eq:scale-x} becomes
\be
\left(2 \Delta +   y_i \partial_{y_i}+  z \partial_z +  z' \partial_{ z'} \right)\left\langle \O(y,z) \O(0,z') \right\rangle  =0 \, ,
\label{scale_position}
\ee
We now multiply this equation by $e^{i \, (\kappa z - \kappa' z')} e^{i \, p  y}$ and integrate in $y$ and in $z,z'>0$. We then try to rewrite the result as a differential operator acting on $G(p,\kappa,\kappa')$. 

We can recognize the first two terms in Eq.~\eqref{scale_position} as those of the scale WI in the absence of a boundary, but for a $(d-1)$-dimensional theory. As before, they give rise to the differential operator
\be
\left(2\Delta-d+1   -  p_i \partial_{p_i} \right) G(p,\kappa,\kappa')\, .
\ee
The last two terms in~\eqref{scale_position} require more care because of possible boundary contributions. For the $z \partial_z$ term we use the integration by parts identity
\be
\int_0^\infty dz \, e^{i\kappa z} \, z\, \partial_z f(z) =  \Big[ e^{i\kappa z} \, z\,  f(z) \Big]^\infty_0  - \left(1+ \kappa \partial_\kappa\right) \int_0^\infty dz \, e^{i\kappa z} \, f(z) ,
\label{eq:byparts}
\ee
with $f(z)= \left\langle \O(y,z) \O(0,z') \right\rangle$. 

We need to understand the behavior of $f(z)$ as $z\to 0$, which we do using the  leading term of the boundary OPE~\eqref{eq:b2b0}.
We will assume that $\Delta-\hat\Delta_1<1$, a condition amply satisfied in the ordinary 3D Ising transition, see Tab.~\ref{tab:3DIsing}. Then, $z\, f(z) \to 0$ as $z\to 0$ and the $z=0$ boundary term in~\eqref{eq:byparts} vanishes. The $z=\infty$ boundary term also vanishes since we are assuming $\Im\kappa>0$. The $z' \partial_{z'}$ term is similar.

Putting everything together, we obtain the following "momentum-space scale WI"
\be
\text{Scale:}\quad\left(2\Delta-d-1 -  p_i \partial_{p_i}  - \kappa \partial_ \kappa -  \kappa' \partial_ {\kappa'} \right) G(p,\kappa,\kappa')=0   \, .
\label{eq:mom0}
\ee

The derivation for the conformal WI is only slightly more involved. We start with the position-space conformal WI~\eqref{eq:scale-x} where we take the index $\mu=i\in\{1,\ldots,d-1\}$ so that the boundary is preserved by the corresponding infinitesimal transformation, and separate variables into $y$ and $z$:
\be
\label{sct_general}
\sum_{a=1,2} \left(2 \Delta y^i_a  +  2 y_a^i (y_a \cdot \partial_{y_a}) +  2 y_a^{i} z_a \partial_{z_a} - (y_a^2+z_a^2) \partial_{ y_a^i}\right)\left\langle \O(y_1,z_1) \O(y_2,z_2) \right\rangle  =0 \, .
\ee
We now set $y_1=y$, $y_2=0$ and obtain:
\be \label{sct_position}
\left(2 \Delta y_{i}  +  2 y_{i} (y \cdot \partial_y)- y^2 \partial_{y_i} +  2 y_{i} \,z \partial_z - z^2 \partial_{y^i}+z'^2 \partial_{ y^i}\right)\left\langle \O(y,z) \O(0,z') \right\rangle  =0 \, .
\ee
(We replaced $\partial_{y_2^i}$ by $-\partial_{y_1^i}$ in the last term, using translation invariance along the boundary.)

We then proceed as for the scale WI. The first three terms in Eq.~\eqref{sct_position} give rise to the $(d-1)$-dimensional differential operator
\be
-i \left(2 (\Delta-d+1) \partial_{p_i}  -  2 \,p_j \partial_{p_j}  \partial_{p_i} + p_i \partial_{p_j}  \partial_{p_j} \right) .
\ee
The term involving $z \partial_z$ is treated by the identity~\eqref{eq:byparts}. Finally, for the term $\propto z^2$ we use
\be
\int_0^\infty dz \, e^{i\kappa z} \, z^2  f(z) = -\partial_\kappa^2 \int_0^\infty dz \, e^{i\kappa z} \, f(z) ,
\ee
and analogously for the one $\propto z'^2$.

Altogether the ``momentum-space conformal WI'' takes the form
\be
\label{eq:mom1}
\text{Conf:}\quad\left(2 (\Delta-d) \partial_{p_i}  -  2 \,p_j \partial_{p_j}  \partial_{p_i} + p_i \partial_{p_j}  \partial_{p_j} -2\kappa \partial_\kappa \partial_{p_i} + p_i \partial_\kappa^2 - p_i \partial_{\kappa'}^2\right) G(p,\kappa,\kappa') =0.
\ee

Differential Eqs.~\eqref{eq:mom0} and~\eqref{eq:mom1} are not directly experimentally useful, because the function $G(p,\kappa,\kappa')$ can be experimentally extracted only for $\kappa'=\kappa^*$. We discuss next how to restrict them to this submanifold.

\subsection{Ward Identity in Half-Space II: Restriction to Physical Values $\kappa'=\kappa^*$}
\label{sec:restriction-to-the-submanifold-kappakappa}
\def\x{u}
\def\y{v}
\def\k{\kappa}
 The theory scattering rate $\mathcal{G}(p,\kappa)$ is the restriction of $G(p,\kappa,\kappa')$ to the two-dimensional real submanifold ${M=\{\kappa'=\kappa^*\} }$ inside $\mathbb{C}^2\ni (\kappa,\kappa')$. We parametrize this submanifold by the real and imaginary parts of $\kappa$:
 \beq
 u_1 = \Re \kappa, \quad u_2=\Im\kappa.
 \eeq 
 So in this section we will think of $\mathcal{G}(p,\kappa)$ as $\mathcal{G}(p,u_1,u_2)$. Starting from the WIs~\eqref{eq:mom0} and~\eqref{eq:mom1}, we would like to derive the differential equations satisfied by $\mathcal{G}$. These differential equations should involve only differential operators acting within the submanifold $M$ on which $\mathcal{G}$ is defined.
 
 Now, if we just restrict equations~\eqref{eq:mom0},~\eqref{eq:mom1}, naively we do not get a closed differential equation for $\mathcal{G}$. In fact, the operators 
\beq 
\kappa \partial_\kappa+\kappa' \partial_{\kappa'},  \kappa \partial_{\kappa}, \partial_\kappa^2-\partial_{\kappa'}^2 
\label{eq:ops}
\eeq
appearing in those equations involve both derivatives tangential to $M$ and orthogonal to~$M$. The latter derivatives cannot be evaluated just knowing $\mathcal{G}$.
 
 There is however one crucial help: the function $G(p,\k,\k')$ is analytic in $\kappa$, $\k'$. Thus it is annihilated by the Cauchy-Riemann operators $\bar\partial_\k$, $\bar\partial_{\k'}$. Hence it suffices to express the operators in~\eqref{eq:ops} on $M$ as operators involving only derivatives along $M$ modulo (linear combinations of products of) such $\bar\partial$ operators. This turns out to be possible (App.~\ref{app:restr}):
 \begin{align}
 \kappa \partial_\kappa+ \kappa' \partial_{\kappa'}& =  (\x_1\partial_{\x_1}+\x_2\partial_{\x_2}) +\ldots 
 \label{eq:op1}\\
2 \kappa \partial_{\kappa}& = (\x_1\partial_{\x_1}+\x_2\partial_{\x_2}) +i (\x_2\partial_{\x_1}-\x_1\partial_{\x_2}) +\ldots 
 \label{eq:op2}\\
 \partial_\kappa^2-  \partial_{\kappa'}^2&=-i  \partial_{\x_1}\partial_{\x_2} +\ldots 
 \label{eq:op3}
 \end{align}
 where $\ldots$ stands for two types of terms: differential operators with coefficients which vanish on $M$, and differential operators involving $\bar\partial_\k$, $\bar\partial_{\k'}$ which give a vanishing result when acting on $G(p,\k,\k')$. Using these relations, we  restrict the WIs~\eqref{eq:mom0},~\eqref{eq:mom1} to $M$, and obtain the WIs for $\mathcal{G}$:
 \begin{align}
 	& \text{Scale: }\quad [2\Delta-d-1 -  p_i \partial_{p_i}  - \x_1\partial_{\x_1}-\x_2\partial_{\x_2} ] \mathcal{G}=0   \, ,
 	\\
    &
   \text{Conf$_1$: }\quad [2 (\Delta-d) \partial_{p_i}  -  2 \,p_j \partial_{p_j}  \partial_{p_i} + p_i \partial_{p_j}  \partial_{p_j} -(\x_1\partial_{\x_1}  +\x_2\partial_{\x_2})\partial_{p_i} ] \mathcal{G}=0 \, ,\\
    &  \text{Conf$_2$: }\quad  [(\x_2\partial_{\x_1} -\x_1\partial_{\x_2})\partial_{p_i} +p_i  \partial_{\x_1}\partial_{\x_2} ] \mathcal{G}  =0  \, .
	\end{align}
In the last two equations we separated the conformal WI into its real and imaginary parts, using the fact that $\mathcal{G}$ is real, as is seen from its definition and is also clear physically from the fact that the scattering rate is real.

The final step is to impose the constraints of rotation invariance along the boundary, which implies that $\mathcal{G}$ only depends on $|p|$. Denoting $r=p^2$, the three WIs become
\begin{align}
& \text{Scale: }\quad [2 \Delta-d -1 -\x_1\partial_{\x_1} -\x_2\partial_{\x_2} -2r \partial_r]\mathcal{G}=0 \, ,	
	\\
& \text{Conf$_1$: }\quad
	[(2 \Delta-d - 3 - \x_1\partial_{\x_1} -\x_2\partial_{\x_2})\partial_r -2 r\partial^2_r ] \mathcal{G}=0 \, ,
\end{align}
and
\beq
\colorbox{lightgray}{$\text{Conf$_2$: }\quad \bigl[2(\x_2\partial_{\x_1}-\x_1\partial_{\x_2})\partial_r+  \partial_{\x_1}\partial_{\x_2}\bigr] \mathcal{G}(p,u_1,u_2)  =0 \, .$}
	\label{eq:C20}
\eeq
Like in the full space case, one conformal WI (Conf$_1$) can be obtained from Scale, by applying $\partial_r$. On the other hand, here in half-space we have an additional conformal WI (Conf$_2$) which is independent from Scale. This Conf$_2$ conformal WI is the main theoretical result of our paper. 

Recall that scale invariance allows to write $\mathcal{G}(p,\kappa)$ in the form~\eqref{eq:Gscale} in terms of a function $g(p, \phi)$. In the current setting where we are using the variable $r=p^2$, that equation becomes (for $d=3$)
\beq
\mathcal{G}(p,\kappa) = |\kappa|^{2\Delta-4} \tilde g(r/|\kappa|^2, \phi)\,,\quad (\kappa = i|\kappa| e^{i\phi}),
\label{eq:phi-param}
\eeq
where the function $\tilde g$ is related to $g$ by $\tilde g(p^2,\phi)=g(p,\phi)$.

We can then rewrite the Conf$_2$ conformal WI in terms of $\tilde g(r,\phi)$ as (for $d=3$)
	\begin{empheq}[box=\colorbox{lightgray}]{equation}
\begin{aligned}
&\Bigl\{2\partial_r\partial_\phi+\cos(2\phi)\bigl[(2\Delta-5)\partial_\phi-2 r \partial_r\partial_\phi \bigr]\\
&\hspace{1cm}+ \frac 12\sin(2\phi)\bigl[4(\Delta-2)(\Delta-3) -\partial_\phi^2+4r^2 \partial_r^2-8(\Delta-3)r \partial_r\bigr]\Bigr\}\tilde g(r,\phi)=0\,.
\label{eq:C2g0}
\end{aligned}
	\end{empheq}
The verification of~\eqref{eq:C20} or~\eqref{eq:C2g0} using an experimentally measured scattering rate would constitute a direct test of conformal invariance of critical fluctuations. See Section~\ref{sec:testing-conformal-invariance}.

\section{Theory: Ward Identity and Small $p$ Expansion}
\label{sec:WIsmall}

Having derived the general consequences of conformal invariance on the theory scattering rate $\mathcal{G}(p,\kappa)$, we consider now how Eq.~\eqref{eq:C2g0} constrains the small $p$ expansion~\eqref{eq:gpphi}, which rewritten in terms of $\tilde g(r,\phi)$ takes the form
\beq
\label{eq:g0_ae}
\tilde g(r,\phi) =  \sum_{i,j=1}^N a_{ij}(\phi) r^{(\hat\Delta_i+\hat\Delta_j)/2-1} + \sum_{k=0}^{k_N-1} e_k(\phi) r^{k}+O(r^{k_N}) \, ,
\eeq
where $N$ is an arbitrary integer counting the number of boundary operators included in the boundary OPE, and $k_N=\lfloor\hat\Delta_{N+1}\rfloor-1$.

In App.~\ref{app:smallp} the coefficients $a_{ij}(\phi)$ are given by explicit integrals, from which various of their properties can be inferred. Importantly for the present discussion, conformal invariance implies selection rules. Namely, the only nonzero $a_{ij}$ correspond to pairs of boundary operators belonging to the same conformal multiplets, and all such pairs have $\hat{\Delta}_j-\hat{\Delta}_i\in 2\mathbb{Z}$. Since we also know that (Eq.~\eqref{eq:aij})
\beq
a_{ij}(\phi) \propto e^{i\phi(\hat\Delta_j-\hat\Delta_i) },
\label{eq:angle}
\eeq 
this restricts the dependence on the phase of the non-integer power part. We will see in a moment how this fact can be given an alternative derivation from the conformal WI. 

The differential operator in the l.h.s.~of~\eqref{eq:C2g0} contains only derivatives in $r$ and integer powers of $r$. It will lead to relations between coefficients of different powers of $r$ in the expansion~\eqref{eq:g0_ae}, with integer spacing. Let us therefore reorganize the small $p$ expansion into families with a leading, possibly non-analytic, scaling in $r$ and analytic corrections~to~it:
\beq
\tilde g(r,\phi) =  \sum_\delta \sum_{k=0}^{k_{N,\delta}} C_{\delta,k}(\phi) r^{\delta-1+k} +O(r^{k_N}) \, ,
\eeq
with $k_{N,\delta}= k_N-\lfloor \delta \rfloor$.
In particular the analytic terms in Eq.~\eqref{eq:g0_ae} correspond to $\delta=1$ (${C_{1,k} =e_k}$), while for the first non-analytic family we have $\delta=\hat{\Delta}_1$ ($C_{\hat{\Delta}_1,0}=a_{11}$). In a conformal theory, all the dimensions $\delta\neq 1$ correspond to scaling dimensions of boundary scalar primaries. 

Before addressing the general case, let us consider some simple consequences of the Conf$_2$ WI~\eqref{eq:C2g0}  for the lowest order terms 
\beq
\label{eq:smallp}
\tilde g(r,\phi) = e_0 (\phi)+ e_1(\phi) r + a_{11}(\phi) r^{\hat\Delta_1-1} +  b_{11} (\phi)r^{\hat\Delta_1} + \dots  \, ,
\eeq
where we defined $b_{11}= C_{\hat{\Delta}_1,1}$.
Start by noticing that the first term in the WI, $2\partial_r\partial_\phi$, is the only one that lowers the degree of $r$. Since $r^{\hat\Delta_1-1}$ corresponds to the leading power in its family and is not constant, it follows that the only way in which the WI can be satisfied is if $\partial_\phi a_{11}=0$. We thus recover the conclusion that $a_{11}$ is a constant. The same considerations hold also for other $C_{\delta,0}$ with $\delta\ne 1$, which satisfy $\partial_\phi C_{\delta,0}=0$.

Consider next the first subleading term $b_{11}$. Using that $\partial_\phi a_{11}=0$, we find 
\beq
\hat\Delta_1 \partial_{\phi} b_{11} +  a_{11} (\hat\Delta_1-\Delta +2)(\hat\Delta_1-\Delta +1) \sin(2\phi)=0 \, .
\eeq
This implies that $b_{11}$ depends on $\phi$ as $\cos(2\phi)$, with a prefactor fixed in terms of $a_{11}$. This is the contribution to $b_{11}$ associated with the boundary descendant $\partial_{\perp}^2 \hat{\mathcal{O}}_1$ with 
dimension $\hat{\Delta}= \hat{\Delta}_1+2$, paired to $\hat{\mathcal{O}}_1$. This is consistent with the angular dependence implied by Eq.~\eqref{eq:angle} for such a pair.\footnote{We get $\cos(2\phi)$ summing over both orderings of the operators within the pair.} By the same result, a constant contribution to $b_{11}$ would be associated to a pair of scalar operators with dimension $\hat{\Delta}= \hat{\Delta}_1+1$. In a conformal theory there is no such operator, since descendants of $ \hat{\mathcal{O}}_1$ with one derivative are not scalars. Therefore, we fix the integration constant to zero and obtain
\beq \label{eq:b11WI}
b_{11} (\phi)= a_{11} \times \dfrac{(\hat\Delta_1-\Delta +2)(\hat\Delta_1-\Delta +1)}{2 \hat\Delta_1} \cos(2\phi)  \, , \qquad a_{11}=const \, .
\eeq
Similarly, we obtain a differential equation relating $e_0$ and $e_1$:
\beq \label{eq:e11WI}
2\partial_\phi e_1+\left( \sin(2\phi)2(2-\Delta)(3-\Delta)+\cos(2\phi)(2\Delta-5)\partial_\phi   -\frac 12 \sin(2\phi)\partial_\phi^2\right) e_0=0 \, .
\eeq
Note that in this case we cannot argue that $\partial_\phi e_0=0$. Indeed we will see below that it has nontrivial dependence on $\phi$ for the 3D ordinary Ising transition.

More generally we can write a differential equation for the coefficients $C_{\delta,k+1}$ and $C_{\delta,k} $:
\beq
\partial_\phi C_{\delta,k+1}= \left(\alpha_{\delta,k} \sin(2\phi) +\beta_{\delta,k}\cos(2\phi)\partial_\phi + \gamma_{\delta,k} \sin(2\phi)\partial_\phi^2\right) C_{\delta,k} \, ,
\eeq
where
\beq
\alpha_{\delta,k} = -\dfrac{(\delta-\Delta+k+1)(\delta-\Delta+k+2)}{\delta+k}, 
\beta_{\delta,k} = \dfrac{2(\delta-\Delta+k)+3}{2(\delta+k)},  \gamma_{\delta,k} = \dfrac{1}{4(\delta+k)} .
\eeq
Together with the constraint $\partial_\phi C_{\delta,0}=0$ for $\delta\neq 1$, this captures the consequences of the Conf$_2$ conformal WI on the small $p$ expansion.

\begin{remark}\label{rem:Gompper}
The prefactor in Eq.~\eqref{eq:b11WI} can be alternatively derived along the lines described in App.~\ref{app:smallp}, using the boundary OPE expansion~\eqref{eq:crossingsy0}, including the first boundary primary and its first scalar descendant. At large $|y|$ we have the following contribution to $G(\xi)$
\beq
G(\xi) \supset \mu_1^2 \left(\dfrac{y^2}{4z z'}\right)^{\Delta -\hat \Delta_1}\left(1-\hat \Delta_1 \frac{z^2 +z'^2}{y^2} +\dots \right) ,  \qquad y^2 \gg z^2,z'^2\, ,
\eeq
from which Eq.~\eqref{eq:b11WI} follows after integration. The relative coefficient of the subleading term can be obtained by imposing conformal invariance of the OPE expansion or, equivalently, by expanding the boundary conformal block \eqref{eq:crossingsy0-block}. 

The same result was previously derived by Ref.~\cite{Gompper1985}, via an argument in the same spirit that made use of the OPE plus conformal invariance. Ref.~\cite{Gompper1985} primarily connected this result to the $\epsilon$-expansion. While a numerical test of this relation will be presented in the next section, an experimental test appears difficult as it requires extracting the subleading coefficient~$b_{11}(\phi)$, which to our knowledge has never been attempted. Our Conf$_2$ conformal WI looks more suitable in this respect. To our knowledge, the full consequences of conformal invariance, as implied by the Conf$_2$ WI, have not been previously intuited or explored.
\end{remark}

\section{Numerical Tests: Ward Identity and Small $p$ Expansion for Ordinary 3D Ising Transition}
\label{sec:num}

In this section we would like to validate on a numerical example the conformal WI and the small $p$ expansion of the theory scattering rate $\mathcal{G}(p,\kappa)$, that we derived in full generality.
We will do so for the ordinary 3D Ising transition in the presence of a plane boundary. This is both theoretically sound, thanks to the accurate knowledge of conformal data available from bootstrap studies~\cite{Chang:2024whx,Reehorst:2021hmp,Gliozzi:2015qsa}, and observationally relevant, since many experimentally accessible systems fall in this universality class, e.g.~\cite{mailander1990near,mailander1990phase,burandt1993near}.
This will also allow us to present some theoretical predictions on the observable grazing scattering rate. Moreover, this numerical study will let us mimic and highlight some of the nuances of experimental measurements.

As already discussed, scale invariance implies that $\mathcal G(p,\kappa) $ can be expressed by Eq.~\eqref{eq:Gscale} in terms of the function $g(p,\phi)$. Setting $\kappa=i e^{i\phi}$ in the expression~\eqref{eq:grazing-sigma} for $\mathcal G(p,\kappa) $, and also using~\eqref{eq:2ptbdry}, the function $g(p,\phi)$ can be written as
\beq
g(p,\phi) = \int_{z,z'>0} dz\, dz'\, \cos[(z'-z)\sin\phi ] e^{-(z+z')\cos\phi} \int d^2y\,  e^{i p y} \frac{1}{[y^2+(z-z')^2]^{\Ds}} G(\xi).
\eeq
We see in particular that the integral is invariant under $\phi\to -\phi$. 

The function $G(\xi)$ defined in Eq.~\eqref{eq:2ptbdry} can be computed to good accuracy from the conformal data of Tab.~\ref{tab:3DIsing}, as reviewed in Sec.~\ref{sec:gxi-for-3d-ordinary-ising-transition}. We rewrite the integral by changing order of integration, and variables to $u=z-z'$ and $ut=z+z'$. We obtain 
\beq
g(p,\phi)= \int d^2y\, e^{ipy} I(y,\phi) \, ,
\label{eq:Iy_integral0}
\eeq
with
\begin{subequations}
\label{eq:Iy_integral}
\begin{align}
I(y,\phi)&=\frac{1}{y^{2\Ds}}  \int_0^\infty  du \frac{\cos(u\sin\phi )}{[1+u^2/y^2]^{\Ds}} R(u,y,\phi)\,,\\ 
R(u,y,\phi) &= u \int_1^\infty dt  \;  e^{-t u \cos\phi}  \;G\left(\frac{y^2/u^2+1}{t^2-1}\right) \, .
\end{align}
\end{subequations}
As detailed in App.~\ref{app:numint}, we determine the leading asymptotic behavior of $I(y,\phi)$ at small and large $y$ 
\beq
I(y,\phi) = 
\begin{cases} 
C_0(\phi)/y^{\alpha}, &y\to 0,\\
C_\infty/y^{\beta}, &y\to\infty,
\end{cases}
\eeq
where
\beq
\begin{split}
&\alpha=2 \Ds-1 \, , \qquad C_{0}(\phi) =\dfrac{1}{ \cos\phi} \times \dfrac{\sqrt{\pi}}{2} \dfrac{\Gamma\left(\Ds-1/2\right)}{\Gamma\left(\Ds\right)} \, , \\
&\beta=2 \hD \, , \qquad \qquad C_{\infty} = \mu_1^2 \, 4^{\hD-\Ds} \, \Gamma(1+\hD-\Ds )^2 \, .
\end{split}
\eeq
Note that $\alpha<2$ while $\beta>2$.
%This asymptotics allows us to extract analytically the coefficient $a_{11}$ which determines the leading non-analytic small $p$ behavior:
%\beq \label{eq:a11analytic}
%a_{11}=  \left(\frac{4\pi\, C_{\infty}}{4^{\hD}}\right)   \frac{ \Gamma(1-\hD) }{\Gamma(\hD)}= \mu_1^2 \left(\frac{4\pi}{4^{\Ds}}\right) \, \frac{ \Gamma (1-\hD) \Gamma(1+\hD-\Ds)^2  }{\Gamma (\hD)} \approx -19.8631 \; ,
%\eeq
%a result we will confirm with the numerical computation.
These asymptotics provide an alternative method to extract analytically the coefficient $a_{11}$ which determines the leading non-analytic small $p$ behavior:
\beq \label{eq:a11analytic}
a_{11}=  \left(\frac{4\pi\, C_{\infty}}{4^{\hD}}\right)   \frac{ \Gamma(1-\hD) }{\Gamma(\hD)}= \mu_1^2 \left(\frac{4\pi}{4^{\Ds}}\right) \, \frac{ \Gamma (1-\hD) \Gamma(1+\hD-\Ds)^2  }{\Gamma (\hD)} \approx -19.8631 \; ,
\eeq
which agrees with the result of Eq.~\eqref{eq:ai} obtained from the boundary OPE expansion.
We will confirm this result with the numerical computation.
\begin{figure}[h]
\centering
\includegraphics[width=0.8\textwidth]{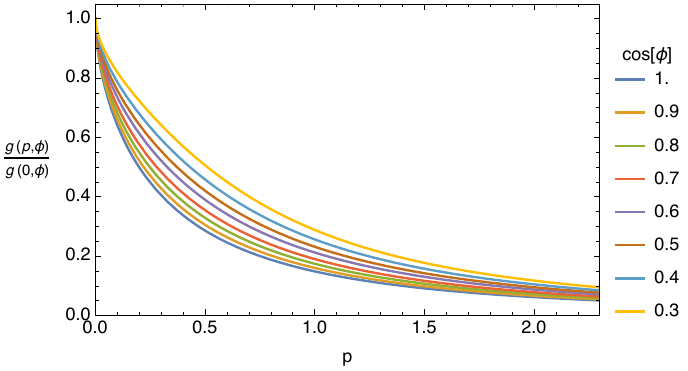}
\caption{The function $g(p,\phi)$ normalized to its value at $p=0$. For illustration purposes we chose values of $\phi$ such that $\cos\phi$ is evenly spaced between $1$ (bottom curve) and $0.3$ (top curve). The normalization at $p=0$ corresponds to $e_0(\phi)$ as reported in Fig.~\ref{fig:e0e1}.\label{fig:gpNormalized}}
\end{figure}

Computing numerically the integrals in Eq.~\eqref{eq:Iy_integral} as described in App.~\ref{app:numint}, we obtain the function $g(p,\phi)$ for a discrete set of values of $\phi$. See Fig.~\ref{fig:gpNormalized} for a representative plot.

\begin{figure}[h]
\centering
\includegraphics[width=0.49\textwidth]{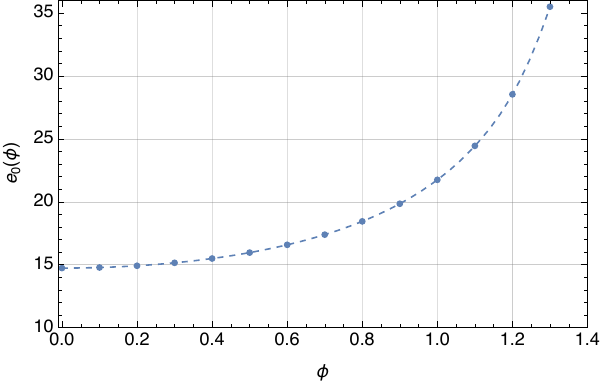}\hfill
\includegraphics[width=0.49\textwidth]{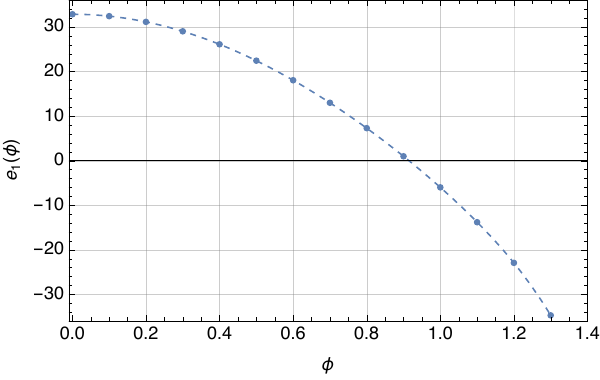}
\caption{First two analytic terms in the small $p$ expansion of $g(p,\phi)$, $e_0(\phi)$ and $e_1(\phi)$, see~\eqref{eq:smallp} for their definition. The coefficients are obtained by direct numerical integration of the function $I(y,\phi)$ for discrete values of $\phi$, as explained in App.~\ref{app:numint}.
\label{fig:e0e1}}
\end{figure}

Using this result we can check the validity of the small $p$ expansion~\eqref{eq:smallp} and extract the value of the coefficients from a fit. For the first two analytic terms, $e_0(\phi)$ and $e_1(\phi)$ we can also determine their value directly from integration of $I(y,\phi)$, as detailed in App.~\ref{app:numint}. The two procedures are in good numerical agreement and we report the numerical values of $e_0(\phi)$ and $e_1(\phi)$ as obtained from direct integration in Fig.~\ref{fig:e0e1}. The intensity of the scattering rate, captured by $e_0$, grows for $\phi\to \pi/2$ as $1/\cos\phi$, in the same way as the penetration depth $1/\kI$. In this limit the process becomes bulk dominated. The $p^2$ coefficient $e_1(\phi)$ has a zero for $\phi \approx 0.915$, which corresponds to $\cos\phi \approx 0.610$. This value is in good agreement with the $\epsilon$-expansion result of Ref.~\cite{gompper1986universal}.\footnote{In the free theory limit (also known as the mean field theory approximation), one can obtain an analytic expression, implying a zero for $\cos\phi = 1/2$, see Eqs.~(5.2)--(5.3) of~\cite{Dietrich1984}.} (See also Remark~\ref{rem:zeta0} below).

\begin{figure}[h]
\centering
\includegraphics[width=0.49\textwidth]{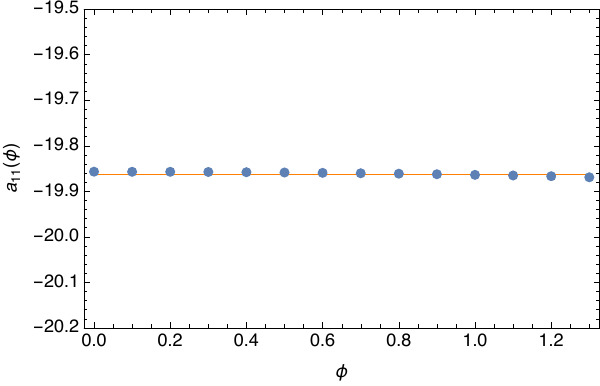}\hfill
\includegraphics[width=0.49\textwidth]{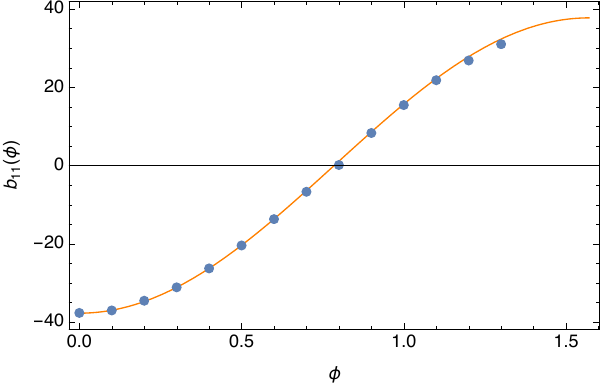}
\caption{First two non-analytic terms in the small $p$ expansion of $g(p,\phi)$, $a_{11}$ and $b_{11}$, see~\eqref{eq:smallp} for their definition. The coefficients are obtained by a numerical fit of the function $g(p,\phi)$ at small~$p$, for discrete values of $\phi$. The results are compared with the analytic predictions of equations~\eqref{eq:b11WI} and~\eqref{eq:a11analytic} (in orange), finding good agreement.  \label{fig:a11-b11}}
\end{figure}

We can similarly extract from a numerical fit the coefficients of the leading and subleading non-analytic terms in the small $p$ expansion~\eqref{eq:smallp}, $a_{11}(\phi)$ and $b_{11}(\phi)$, see Fig.~\ref{fig:a11-b11}. For the former we obtain values consistent with a constant independent of $\phi$, as expected, and in good numerical agreement with the analytic prediction~\eqref{eq:a11analytic}. For the latter, we find an oscillating behavior consistent with $\cos(2\phi)$, with zero average and amplitude fixed by $a_{11}$ as predicted by Eq.~\eqref{eq:b11WI}. 

\begin{figure}
	\centering
	\vspace{5pt}
	\includegraphics[width=0.7\textwidth]{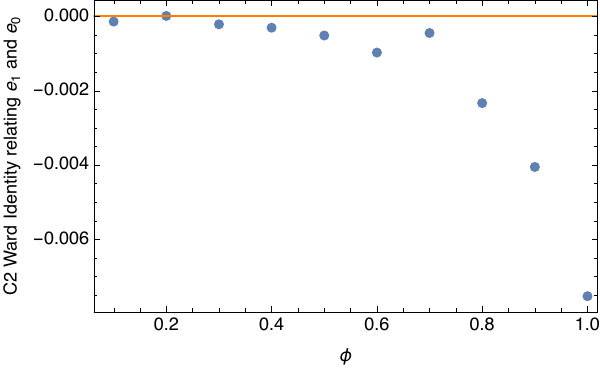}
	\caption{Numerical test of the small $p$ conformal Ward identity~\eqref{eq:e11WI}, as a function of $\phi$. We plot the value of the l.h.s. of Eq.~\eqref{eq:e11WI} normalized by the value of its first term. \label{fig:e1WI}}
\end{figure}

We can use these results to test the conformal WI. Let us start from the small $p$ expansion and check Eq.~\eqref{eq:e11WI}. With the numerical accuracy of our computation, the WI is satisfied to the level of $\sim 10^{-3}$ for angles $\phi<1$, with the accuracy reaching the $10^{-4}$ level for small $\phi$ and degrading for large values of $\phi$ as shown in Fig.~\ref{fig:e1WI}. 
This test of the WI requires first extracting the small $p$ expansion coefficients $e_0(\phi)$ and $e_1(\phi)$ as a function of~$\phi$, a procedure that we carried out in this work in a numerical setting, but which might be challenging in an experimental setting due to finite experimental accuracy.\footnote{In practice one will have to do a joint fit for four quantities $e_0(\phi)$, $e_1(\phi)$, $a_{11}$ and $b_{11}(\phi)$.}

\begin{figure}
\centering
\includegraphics[width=0.54\textwidth]{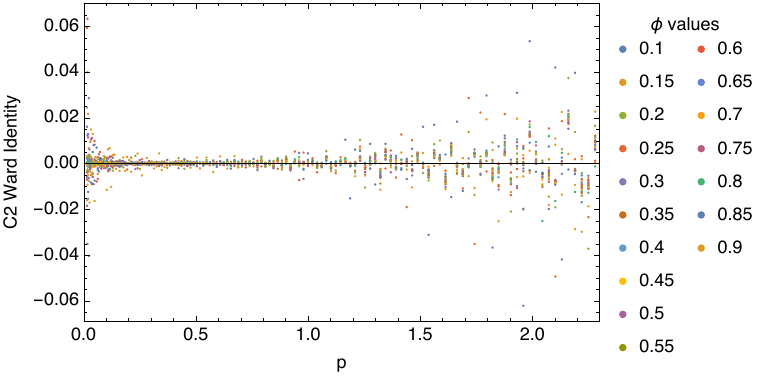} \hfill
\includegraphics[width=0.41\textwidth]{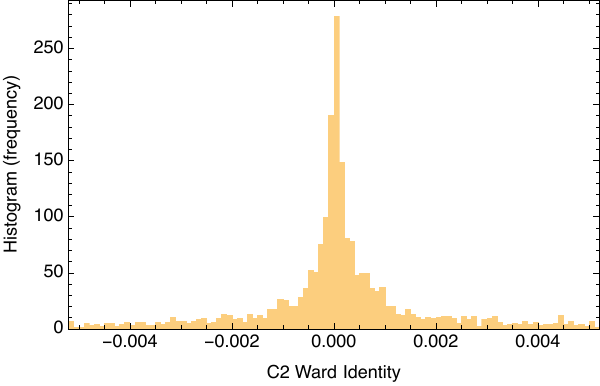}
\caption{Numerical test of the conformal Ward identity~\eqref{eq:C2g0}, as a function of $0.01<p<2.3$ for discrete values of $\phi$. We give the value of the l.h.s. of Eq.~\eqref{eq:C2g0} normalized by the value of its first term. The complete results are shown in the left panel as a scatter plot, clustered around $0$ up to numerical errors. The results are combined into a histogram in the right panel. We find a distribution with mean $1.6 \times 10^{-4}$ and standard deviation $5.9 \times 10^{-3}$. \label{fig:WItest}}
\end{figure}

An alternative approach is to test the Conf$_2$ Ward identity in the form~\eqref{eq:C2g0} directly at finite momentum ($p$ or, equivalently, $r=p^2$), in the whole range of available momenta. In our numerical test we do so by discretizing the derivatives in $\phi$ and $r$ and evaluating them on the discrete set of data points computed for $0.1<\phi<0.9$ with a constant spacing of $\Delta\phi=0.05$. We evaluate discrete derivatives using a five-point stencil, and as a consequence the data used extend from $\phi=0$ to $\phi=1$. The momentum is taken in the range $0.01\leq p \leq 2.3$, with~$\sqrt{p}$ uniformly spaced in steps of $0.01$.\footnote{The natural variable is the physical momentum $p$, used in all our plots in line with experimental works. In studying the Ward identity however, it is convenient to adopt $r$ as momentum variable due to rotational invariance. The numerical computations are performed with $\sqrt{p}$ evenly spaced to ensure an approximately uniform sampling of the cross-section, which behaves roughly as $|p|^{0.552}$ at small $p$. The discrete derivatives are first computed with respect to $\sqrt{p}$, from which the $r$ derivatives are obtained by the chain~rule.}
The complete results are shown in the left panel of Fig.~\ref{fig:WItest} as a scatter plot. We see that the results cluster around $0$ for all momenta, with larger numerical noise at very small or larger momenta. In order to quantify the agreement with the expected validity of the WI, we can combine the data by marginalizing over $\phi$ and $p$, obtaining the histogram shown in the right panel of Fig.~\ref{fig:WItest}. As expected, the data have a bell-shaped distribution centered around zero with mean $1.6 \times 10^{-4}$ and standard deviation $5.9 \times 10^{-3}$. We take this as a convincing numerical test of conformal invariance for the numerically computed observable $\mathcal{G}(p,\kappa)$.

Needless to say, we could do the integrals more carefully. Also, the l.h.s.~of the conformal WI could be evaluated more accurately than with a five-point stencil. 
The numerical test would then agree with zero with even smaller residuals. 
We do not do this because that is not the point. In the next section we will connect with a possible \emph{experimental} test of the conformal WI. Any such test will have experimental errors, and the result of substituting the measured scattering cross-section into our conformal WI will not be exactly zero either. Our imperfect numerical plots should suggest what future experimental plots may look like, even if the source of the error will not be exactly the same.

\section{Experimental Protocol: How to Test Conformal Invariance}\label{sec:testing-conformal-invariance}

Having validated the Conf$_2$ conformal Ward identity, Eqs.~\eqref{eq:C20} and~\eqref{eq:C2g0}, and presented numerical predictions for the scattering rate of the ordinary 3D Ising transition, we proceed in this section with a discussion of how such a test can be achieved experimentally with grazing scattering of X-rays (GISAXS) or neutrons (GISANS). We focus on X-rays in the main text, see Remark \ref{rem:neutrons-exp} for neutrons.

An experimental test of conformal invariance should proceed as follows: 
\begin{enumerate}
	\item
	At $T=T_c$, measure the (exclusive) grazing scattering cross-section in a range of scattering momenta $p$ and in a range of $\kappa$;
	\item
	Correcting for the Fresnel factor, extract the ``theory scattering rate'' $\mathcal{G}(p,\kappa)$, which is the Fourier-Laplace transform of the two-point function of the order parameter, see~\eqref{eq:grazing-sigma};
	\item
	Act on $\mathcal{G}(p,\kappa)$ with the differential operator~\eqref{eq:C20}. Alternatively, compute $\tilde g(r,\phi)$ by rescaling the momentum variable as in~\eqref{eq:phi-param}; then act on it with~\eqref{eq:C2g0}. If the result is zero within experimental errors, this proves conformal invariance.
	\end{enumerate}

Some comments are in order. First of all, the differential equation gives trivially zero for $\kappa$ purely imaginary. For a nontrivial test both the real and imaginary parts of $\kappa$ must be nonzero, equivalently $\phi\ne0$. This means that (see Eq.~\eqref{eq:kappa}):
\beq
\text{one of the angles $\alpha_i,\alpha_f$ should be below $\alpha_c$, one above.}
\label{eq:belowabove}
\eeq
Second, Fresnel factors are known trigonometric functions of $\alpha_i,\alpha_f,\alpha_c$, see e.g.~\cite[App.~A]{Dietrich1984}.

\begin{figure}
	\centering
	\includegraphics[width=0.54\textwidth]{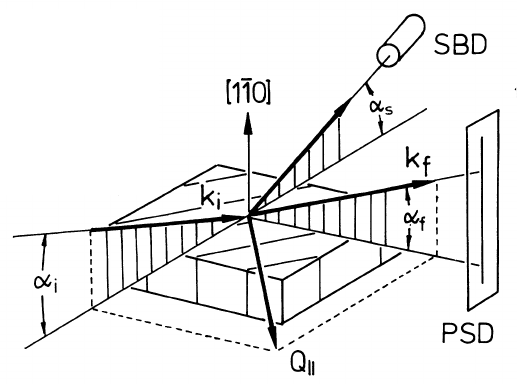} 
	\caption{Experimental setup of Mail\"ander et al., figure from~\cite{mailander1990near,mailander1990phase}. SBD is the specular beam detector and PSD is the position-sensitive detector. \label{fig:exp-setup}}
\end{figure}

Let us discuss the grazing scattering experiment by Mail\"ander et al.~\cite{mailander1990near,mailander1990phase}, see Fig.~\ref{fig:exp-setup}. The authors verified the cuspy behavior of the cross-section at small $p$, Eq.~\eqref{eq:asymp0phi}, see Fig.~\ref{fig:Mailander}. As we will see, their setup was well suited for testing conformal invariance. It looks like our test could have been done already back in 1990.

Mail\"ander et al. studied grazing scattering of X-rays on an Fe$_3$Al alloy (actually 29 at.~\%~Al) at its \emph{Fm3m} to \emph{Pm3m} continuous bulk phase transition temperature ${T_c\sim 800 K}$~\cite{PhysRevLett.22.517}. They used X-rays of wavelength $\lambda=1.8$~\AA, resulting in $\alpha_c=7.2$~mrad. The incidence angle was $\alpha_i=0.95~\alpha_c$.\footnote{There are discrepancies between~\cite{mailander1990phase} and~\cite{mailander1990near} in the reported values of $\alpha_c$, penetration length $\Lambda$ and the range of $p$, although the cross-section data points look the same. Here we use the values from~\cite{mailander1990near}.} Using~\eqref{eq:kappa} and $K=2\pi/\lambda$ we obtain
\beq
\Im \kappa=K(\sin^2\alpha_i -\sin^2 \alpha_c)^{1/2}=(127.4\text{ \AA})^{-1},
\ee
close to the penetration length $\Lambda=(\Im\kappa)^{-1}=125$~\AA~reported in~\cite{mailander1990near}. (Our $\kappa$ is $Q_z$ in~\cite{mailander1990near}.)

On the other hand, $\alpha_f$ was varied in the range 1.2--1.5~$\alpha_c$. We see that condition~\eqref{eq:belowabove} was satisfied in their experiment. Consequently
\beq
\Re \kappa=K(\sin^2\alpha_f -\sin^2 \alpha_c)^{1/2}
\ee
varied between $(60\text{ \AA})^{-1}$ and $(35\text{ \AA})^{-1}$. The experiment was performed at varying~$|\kappa|$ and~$\phi$. In fact, their main plot Fig.~\ref{fig:Mailander} was obtained by fixing the momentum transfer $|p|$ (denoted $q$ or $|q_{\|}|$ in~\cite{mailander1990near,mailander1990phase}) and integrating the signal over the mentioned range of $\alpha_f$. This was convenient given the geometry of their experimental apparatus; see the position of PSD in~\cite{mailander1990near}, see Fig.~\ref{fig:exp-setup}.

Because of the integration over $\alpha_f$ and variation of $|\kappa|$, it is not possible to easily match their Fig.~\ref{fig:Mailander} with any of the curves in our Fig.~\ref{fig:gpNormalized}, which presumed fixed $|\kappa|$ and $\phi$. Also the Fresnel factor, which depends on $\alpha_f$, gets in the way of this comparison. 

In the future, the experiment of~\cite{mailander1990near} (or a similar one~\cite{burandt1993near}) should be repeated for each~$\alpha_f$ separately.\footnote{We would also like to mention X-ray diffraction experiments of surface criticality~\cite{PhysRevLett.62.1294,PhysRevLett.63.2578}. In those systems, only the 2D surface was critical while the 3D bulk was non-critical. One could perhaps test conformal invariance by introducing a one-dimensional edge into those setups.}  This should be possible; indeed already in 1990 angle-resolved measurements were performed, albeit only at $p=0$ --- see the total intensity shown in the inset of Fig.~\ref{fig:Mailander}. One should also correct for the Fresnel factor before comparing with our theory predictions and checking the conformal WI.

\begin{remark} \label{rem:zeta0}
Ref.~\cite{mailander1990near} (but not~\cite{mailander1990phase}) reports that it is convenient to work at $\cos\phi=\Im \kappa/|\kappa|\sim 0.6$ to remove a correction term to the small $p$ scaling behavior. This matches nicely with what we presented in Section~\ref{sec:num}: the coefficient of the $p^2$ correction term vanishes for $\cos\phi\approx 0.61$. But did Ref.~\cite{mailander1990near} actually work at this special value of $\cos\phi$? Their $\alpha_i,\alpha_f,\alpha_c$ seem to correspond instead to $\cos\phi=\text{0.45--0.27}$.
\end{remark}

\begin{remark} \label{rem:neutrons-exp}
Grazing angle surface neutron scattering (GISANS) can be used to study boundary critical phenomena in magnetic materials, see Remark \ref{rem:neutron}. The main challenges are due to the weak scattering cross-section and small angles between the specular beam and exit beam. For a recent review of experimental progress see~\cite{kohler2025recent} (see also~\cite{als1994principles,muller2013grazing} for earlier reports). We are not aware of neutron experiments verifying the cuspy behavior of the critical scattering rate. Therefore, it is not immediately clear to us if the test of conformal invariance with neutrons is feasible or not with the current techniques. This question deserves a separate study.
\end{remark}
\flushbottom

\section{Conclusions}
\label{sec:conclusions}

The most direct experimental test of conformal invariance would result from testing the kinematics of a correlation function which is constrained by conformal invariance more than by scale invariance alone. Here we argued that a two-point function in the presence of a flat boundary is very much suitable for such a test. The conformal kinematic constraints can be tested by performing grazing-angle surface scattering, where the probe wavefunction exponentially decreases inside the sample, with controllable penetration length. For experimental convenience, we expressed the constraints of conformal invariance as a partial differential equation (``conformal Ward identity'') acting on the measured exclusive cross-section (after correcting for the Fresnel factor) in the space of momentum transfer and scattering angle. This eliminates the need for reconstructing the two-point function in position space.

Our work provides a concrete, feasible way to test conformal invariance in critical phenomena using standard X-ray scattering techniques. Grazing scattering experiments of X-rays for boundary critical phenomena were already performed more than 30 years ago. While published data~\cite{mailander1990near,mailander1990phase,burandt1993near} is not enough to check our conformal Ward identity, it seems that only a minor upgrade of these experiments is needed to achieve this. We encourage joint efforts between experimentalists and theorists to refine and implement such a test. As for neutron scattering, further study is needed to assess their feasibility (see Remark \ref{rem:neutrons-exp}).

While the theoretical results obtained in this work, i.e. the momentum space conformal WI and the small $p$ expansion of the scattering rate, are general, the numerical tests and experimental outlook in our analysis focused on the 3D ordinary Ising universality class.  This is because it corresponds to a widely experimentally studied universality class, and in particular the one realized in prior grazing scattering experiments on surface critical phenomena. In the future, it would be interesting to see if our methods could be fruitfully extended to other systems (e.g., different universality classes, lower dimensions, or other boundary conditions).

\acknowledgments

SR thanks Aleix Gimenez-Grau for an initial remark that two-point functions in the presence of a boundary could be used to test conformal invariance. AP thanks Nikolay Ebel and Giovanni Rizi for useful comments. We are grateful to Marco Meineri for the very useful feedback on a draft of this paper. SR~is partially supported by the Simons Collaboration on the Probabilistic Paths to Quantum Field Theory (award SFI-MPS-PP-00012621-16). The work of AP is supported by the Huawei Young Talents Program at IHES.

\section*{Data availability statement}

The {\tt Mathematica} notebook used to produce the figures in Section~\ref{sec:num} (based on App.~\ref{app:numint}) is available from the authors upon request. 

 \appendix

\section{Scaling Laws Near Criticality}
\label{app:scaling}

In the main text we discussed the behavior of the scattering rate at $T=T_c$. For completeness, in this appendix we review the scaling laws close to the critical point, as a function of temperature, for $T\sim T_c$. We focus on the case of antiferromagnetic-type transitions, such as in binary alloys, to highlight some of their features. 

In an antiferromagnetic-type transition, close to criticality, the potential $V$ in the ordered phase has a periodic component
\beq
V(x)=const.e^{-i Q_0 x}\mathcal{O}(x)+\ldots,
\eeq 
where $Q_0$ is a nontrivial momentum inside the Brillouin zone. The potential $V$ could be associated, for instance, to the atomic density, and its ordered component to a binary AB order. This gives rise to a peak in the probe scattering cross-section (e.g. X-rays), when the momentum transfer $Q$ is close to $Q_0$: $Q=Q_0+P$ with $P$ small. The intensity of this ``superlattice Bragg peak'' goes to zero as the phase transition is approached from the ordered phase. In addition to the superlattice Bragg peak associated to the ordered component, $\langle \mathcal{O}(z) \rangle$, there is an additional component associated to critical fluctuations, $\langle \mathcal O(y,z) \mathcal O(0,z') \rangle$. This component, known as critical diffuse scattering, persists even at the critical point and is the main focus of this work. We will discuss its small $p$ scaling in App.~\ref{app:smallp} and focus here, instead, on scaling laws as a function of the temperature $T$. For definiteness, assume that we only vary $T$ and the system is ordered for $T<T_c$. Moreover, define $t=(T-T_c)/T$. Then the intensity of the superlattice Bragg peak is controlled by the square of the order parameter and scales as
\beq
I_B \sim \vert t \vert^{2\beta},\quad \beta=\Delta/(3-\Delta_1),
\eeq
where $\Delta$ is the scaling dimension of $\mathcal{O}$ and $\Delta_1$ is the scaling dimension of the leading uncharged scalar operator (for the case of the 3D Ising theory, see Tab.~\ref{tab:3DIsing}). Furthermore, at fixed small $t\neq 0$, the system has finite correlation length~$\xi$. This can be measured from the width of the diffuse scattering rate as a function of~$p$, e.g. in the disordered phase ($t>0$). The correlation length diverges at the critical point as
\beq
\xi \sim \vert t \vert^{-\nu} \, ,\quad \nu=1/(3-\Delta_1) \, .
\eeq

In the presence of a boundary, we assume as in the main text that $Q_0$ is parallel to the boundary $z=0$, and $q= Q_0+p$ is the momentum transfer parallel to the boundary. The intensity of the superlattice Bragg peak at $p = 0$ is controlled by the order parameter
\be
\langle \mathcal{O}(z) \rangle \sim \vert t \vert^{\beta} \,  f\left(\dfrac{z}{\xi}\right)\, , \qquad {\rm with} \qquad  f\left(\dfrac{z}{\xi}\right) \sim \left(\dfrac{z}{\xi}\right)^\frac{\beta_1-\beta}{\nu} \quad {\rm for} \quad z/\xi \to 0 \, .
\ee
For large $z$ the order parameter should tend to its infinite-space value and we have  $f\to 1$.
The small $z$ behavior can be derived from the bulk-to-boundary OPE:
\beq
\langle\O(y,z)\rangle \propto \frac{1}{z^{\Delta-\hat\Delta_1}} \langle \hat{\O}_1(y) \rangle \sim \frac 1{z^{\Delta-\hat{\Delta}_1}} t^{\beta_1}  \;, \quad \beta_1 = \hat{\Delta}_1/(3-\Delta_1) \,  .
\eeq
The intensity of the superlattice Bragg peak scales like
\be \label{eq:IBgrazing}
I_B \propto \Bigl |\int_{z>0} \langle \mathcal{O}(z) \rangle e^{i \kappa z} dz \Bigr |^2  \sim  \vert t \vert^{2\beta_1} \,  \vert \kappa \vert^{-2\mu} \, ,
\ee
with $\mu= 1+ (\beta_1-\beta)/\nu$. The proportionality factors carry a dependence on $\kappa$ through the Fresnel transmission coefficients, and possibly the atomic form factors, see also footnote~\ref{atomicFF}.

\section{Small $p$ Expansion of Scattering Rate at Criticality}
\label{app:smallp}

We present here a detailed derivation of the small $p$ expansion for theoretical scattering rate ($T=T_c$) in half-space, Eq.~\eqref{eq:gpphi}. This is an asymptotic expansion in analytic (integer powers of $p^2$) plus non-analytic (non-integer powers of $p$) terms. Non-analytic powers and their coefficients are fixed from the boundary OPE expansion, while analytic terms probe the shape of two-point functions at all distances. 

The theory scattering rate $\mathcal G(p,\kappa) $ is defined in terms of the field theory two-point function in half-space, Eq.~\eqref{eq:grazing-sigma}.
Recall that we parametrized $\kappa =i  | \kappa |e^{i\phi } $ and we rescaled momenta by $| \kappa |$ writing the function $\mathcal G(p,\kappa) $ in terms of $g(p,\phi)$, Eq.~\eqref{eq:Gscale}. 

Let us study first the case of $\kappa$ purely imaginary, i.e. $\phi=0$. Using the symmetry under $y\to -y$, we write $g(p):=g(p,0)$ as 
\beq
g(p) =\int_{z,z'>0} dz\, dz'\,  e^{-z -z'} \int d^2y \cos(p y) \langle \mathcal O(y,z) \mathcal O(0,z') \rangle\,.
\label{eq:g}
\eeq
We start with the simple observation that $g(p)$ is a real nonnegative function of $|p|$ bounded by its value at $p=0$:
\beq
0\le g(p) \le g(0) <\infty\,.
\label{eq:g0}
\eeq
Reality is obvious from~\eqref{eq:g}. Non-negativity is physically expected since it expresses a cross-section; mathematically it can be shown using the spectral decomposition of the two-point function. The bound $g(p) \le g(0)$ is clear since $|\cos(p y)|\le 1$. Let us check that $g(0)$ is finite. We need to check, first, that the $y$-integral converges:
\beq
J(z,z'):=\int d^2y \langle \mathcal O(y,z) \mathcal O(0,z') \rangle <\infty,
\label{eq:intI}
\eeq
and second, that once we plug the result into~\eqref{eq:g} the integral in $z,z'$ converges near zero. (Convergence near infinity is not an issue because of the exponential suppression.)

Let us do a detailed check, since similar techniques apply in more general considerations below. We have the distance between the two operators $l=\sqrt{y^2+(z-z')^2}$ and their distances from the boundary $z,z'$. If they are all comparable, $l\sim z\sim z'$, the correlator is~$\sim l^{-2\Delta}$ by dimensional reasons. If one is smaller than the others, we estimate by the OPE, either bulk or boundary. The resulting estimates are 
\beq
\langle \mathcal O(y,z) \mathcal O(0,z')\rangle \sim 
\begin{cases} 
l^{-2\Delta}, & l\sim z\sim z'\\
l^{-2\Delta}, & l\ll z \sim z'\\
(zz')^{\hat{\Delta}_1-\Delta} l^{-2\hat\Delta_1}, &  l\gg \max(z, z') \\
z^{\hat{\Delta}_1-\Delta} (z')^{-\hat\Delta_1-\Delta}, & z\ll l\sim z'\\
(z')^{\hat{\Delta}_1-\Delta} z^{-\hat\Delta_1-\Delta}, & z'\ll l\sim z
\end{cases}
\label{eq:2pt-est}
\eeq
The second line is by the bulk OPE. The third line is by using the leading boundary OPE
\beq
\O(y,z) = (2z)^{\hat{\Delta}_1-\Delta} \hat{\O}_1(y) +\ldots\, .
\eeq
for both operator insertions (here we set the OPE coefficient to 1 to simplify the notation). In the fourth line we use the boundary OPE for $\O(y,z)$ and estimate the resulting two-point function $\langle\hat{\O}_1(y) \O(0,z')\rangle$ by its natural scaling. Interchanging $z$ and $z'$ we get the last line. Note that these estimates agree parametrically along the common boundaries. 

Recall the ordinary 3D Ising transition dimensions given in Tab.~\ref{tab:3DIsing}. In the estimates below it will be sufficient to assume more generally 
\beq
\Delta<1, \; \hat\Delta_1>1.
\eeq
Using~\eqref{eq:2pt-est}, it is easy to estimate the integral~\eqref{eq:intI}:
\beq
J(z,z')\sim \begin{cases} (z')^{2-2\Delta} \, , & z\sim z'\\
z^{\hat\Delta_1-\Delta} (z')^{2-\hat\Delta_1-\Delta} \, , & z\ll z'
\end{cases}
\eeq
and similarly for $z\gg z'$. These estimates do imply that the $z,z'$ integral in~\eqref{eq:g} converges. This proves that $g(0)$ is finite.

Next we are interested in a small $p$ expansion of $g(p)$. If we Taylor-expand $\cos(p y)$ under the integral sign, then  already at order $p^2$ the integrand goes as $y^{2-2\hat\Delta_1}$ at large $y$, giving a divergent $y$ integral for the 3D Ising $\hat\Delta_1$. Thus, although we have seen that $g(0)$ is finite, its $p$ derivatives at the origin are infinite, i.e.~the function is non-analytic in $p$. 

We will extract the non-analytic terms systematically by the boundary OPE~\eqref{eq:b2b0}, including higher order terms.
The boundary OPE should be valid, at least in an asymptotic sense, even without assuming conformal invariance. In a conformally invariant theory, some operators $\hat{\O}_i$ would be boundary primaries and some scalar descendants, with OPE coefficients $\mu_i$ related to those of the primaries. The two-point function normalization factors $\mathcal{N}_{ij}$ in Eq.~\eqref{eq:Ay} below would then vanish except for operators in the same conformal multiplet.  

Using the first $N$ boundary OPE terms we get 
\beq
\langle \O(y,z) \O(0,z') \rangle = A(y,z,z') + (zz')^{-\Delta} O( (zz'/|y|^2)^{\hat\Delta_{N+1}}) \; ,
\eeq
where
\beq
A(y,z,z') =  \sum_{i,j=1}^N \mu_i\mu_j (2z)^{\hat{\Delta}_i-\Delta} (2z')^{\hat{\Delta}_j-\Delta} \frac{\mathcal{N}_{ij} }{|y|^{\hat\Delta_i+\hat\Delta_j}} \; ,
\label{eq:Ay}
\eeq
and $ \hat\Delta_{N+1}$ in the error term is the first non-included dimension. In a reflection positive theory, such as 3D Ising, the diagonal entries of the $\mathcal{N}_{ij}$ matrix are non-negative.

This boundary OPE gives a good approximation for $y\gg \max(z,z')$. Still, our strategy is to do the $y$ integral over the whole $\mathbb{R}^2$, first for $A(y,z,z')$ and then for the remainder $E(y,z,z'):=\langle \O(y,z) \O(0,z') \rangle - A(y,z,z')$. Of course, in the second integral $E(y,z,z')$ is small only for large $y$. This will suffice for our purposes.

When we integrate $e^{ip y} A(y,z,z')$ in $y$, we encounter divergences at $y=0$. We regulate them in a scale-invariant way, so that $\frac{1}{|y|^{\hat\Delta_i+\hat\Delta_j}}$ is a scale-invariant distribution in $y$. Doing the Fourier transform and the $z,z'$ integral, we obtain the first part of $g(p)$:
\beq
g_A(p) = \sum_{i,j=1}^N a_{ij}  p^{\hat\Delta_i+\hat\Delta_j-2},
\eeq
where
\begin{subequations}
 \label{eq:ai}
\begin{align}
 a_{ij} &= \mu_i\mu_j \mathcal{N}_{ij} 2 ^{2-2\Delta}\pi \frac{\Gamma(1-\frac 12(\hat\Delta_i+\hat\Delta_j))}{\Gamma(\frac 12(\hat\Delta_i+\hat\Delta_j))} Y(\Delta_i,\Delta_j,\Delta) \, ,\\
 Y(\Delta_i,\Delta_j,\Delta)&:=\int_{z,z'>0} dz\, dz' e^{-z-z'} z^{\hat{\Delta}_i-\Delta} (z')^{\hat{\Delta}_j-\Delta} =\Gamma(\hat{\Delta}_i-\Delta+1) \Gamma(\hat{\Delta}_j-\Delta+1) \,.
 \end{align}
 \end{subequations}
 Note in particular that for the ordinary 3D Ising transition value of $\hat\Delta_1$ we have $a_{ 11}<0$, the result mentioned in Sec.~\ref{sec:half-space}, also needed for consistency with~\eqref{eq:g0}.
 
The second part of $g(p)$ is given by
\beq
g_E(p) =  \int_{z,z'>0} dz \,dz' e^{-z-z'} \int d^2 y \cos(py) [\langle \O(y,z) \O(0,z') \rangle - A(y,z,z')]\,.
\label{eq:gE}
\eeq
We will show that this function has a Taylor expansion of the form:
\beq
g_E(p)=\sum_{k=0}^{k_N-1} e_k (p^2)^{k}+O((p^2)^{k_N})\,,
\label{eq:gE_Taylor}
\eeq
where 
\beq
k_N=\lfloor\hat\Delta_{N+1}\rfloor-1.
\label{eq:kN}
\eeq
Let $T_{2k_N-2}(py)$ be the Taylor polynomial of $\cos(py)$ of order $2k_N-2$. Then, for any $py$
\beq
\cos(py) = T_{2k_N-2}(py)+R(py),\quad |R(py)|\le C (py)^{2k_N}\,.
\eeq
For $k_N$ as above we have $2\hat\Delta_{N+1}-2k_N>2$
and so all terms obtained by Taylor-expanding $\cos(py)$, including the error term, are convergent at large $y$. One can check that the remaining integrals in $z,z'$ are also convergent. This proves~\eqref{eq:gE_Taylor}.

We thus showed that $g(p)$ can be represented as a sum of two terms, one with integer and one with generally non-integer powers of $p^2$:
\beq
g(p) =  \sum_{i,j=1}^N a_{ij} p^{\hat\Delta_i+\hat\Delta_j-2} + \sum_{k=0}^{k_N-1} e_k (p^2)^{k}+O((p^2)^{k_N}),
\label{eq:gp-final}
\eeq
which is precisely~\eqref{eq:gpphi} for the considered case $\phi=0$.
Here $N$ is arbitrary and $k_N$ is given by~\eqref{eq:kN}. The coefficients $a_{ij}$ have a simple universal expression~\eqref{eq:ai}. On the other hand, $e_k$ are more complicated, obtained by Taylor-expanding $\cos(py)$ in~\eqref{eq:gE} and computing the $y,z,z'$ integrals. They can be computed if the exact two-point function is known. In generic scale invariant  theories, it is hard to say anything specific about them except that they are finite, for $k$ in the given range.\footnote{As a sanity check, once sufficiently many boundary OPE terms are subtracted so that a given $e_k$ becomes finite, subtracting further terms does not change the result.} (But see the comment at the very end of this appendix.)

We can increase $N$, and then $k_N$ will also get larger. However this will not necessarily improve the accuracy of~\eqref{eq:gp-final} for a given finite $p$, since we do not a priori control the size of the coefficients nor of the $O((p^2)^{k_N})$ error term. We should view~\eqref{eq:gp-final} as an asymptotic expansion rather than as a convergent one.

Finally, consider the general case $\phi\ne 0$.
We still get an expansion in non-integer and integer powers of $p^2$, by an essentially identical argument.  
For the non-integer power part of $g(p,\phi)$, the coefficients $a_{ij}(\phi)$ will be obtained by replacing $Y(\Delta_i,\Delta_j,\Delta)$ in~\eqref{eq:ai} with
\beq
\int_{z,z'>0} dz dz' e^{i\kappa z-i\kappa^* z'} z^{\hat{\Delta}_i-\Delta} (z')^{\hat{\Delta}_j-\Delta} \,,
\label{eq:ai-kappa}
\eeq
where $\kappa=ie^{i\phi}$. We can now rotate the integration contours and reduce to the purely imaginary $\kappa=i$, this integral being equal
\beq
\label{eq:aij}
 e^{i\phi(\hat{\Delta}_j-\hat{\Delta}_i)} Y(\Delta_i,\Delta_j,\Delta) \, .
\eeq
The dependence on the phase cancels for the diagonal terms $i=j$, in particular for the leading term $i=j=1$. In a conformally invariant theory, non-diagonal $a_{ij}$ will vanish for fields in different conformal multiplets because of $\mathcal{N}_{ij}=0$. The remaining nonzero terms will then have $\hat{\Delta}_j-\hat{\Delta}_i\in 2\mathbb{Z}$. This restricts the dependence on the phase of the non-integer power part, in conformal theories. 

As to the coefficients $e_k$ of the integer-power part of $g(p,\phi)$, including $e_0$, their dependence on $\phi$ is more complicated. In a CFT, there will be a differential equation that relates $e_k$ and $e_{k+1}$ (see Sec.~\ref{sec:WIsmall}).

\section{Restrictable Operators}
 \label{app:restr}
In this appendix we explain~\eqref{eq:op1}--\eqref{eq:op3}. Denote $\kappa_1=\kappa$, $\kappa_2=\kappa'$ and let $s_a,t_a$ be the real/imaginary parts of $\kappa_a$. A basis of first-order differential operators on $\mathbb{C}^2$ consists of
\be
\partial_a, \bar\partial_a =\frac 12(\partial_{s_a}\mp i\partial_{t_a})\,.
\ee
The operators $\partial_a$ appear in the WIs for $G(p,\kappa,\kappa')$, while operators $\bar\partial_a$ can be added at will since they annihilate the analytic function $G(p,\kappa,\kappa')$.

Our submanifold $M$ is defined by $\kappa_1 = \kappa_2^*$, i.e. $s_1=s_2$, $t_1=-t_2$. We will work in real coordinates: 
\beq
\x_1 = \frac12(s_1 + s_2) \,, \qquad \x_2 = \frac 12(t_1-t_2)\,, \qquad \y_1 = s_1-s_2 \,,\qquad  \y_2 = t_1+t_2 \,, 
\eeq
In these coordinates $M$ corresponds to $\y_1 = \y_2=0$. On $M$ we have $\x_1=\Re \kappa$, $\x_2=\Im \kappa$ as defined in the main text. 

We express all differential operators in terms of $\partial_{\x_1},\partial_{\x_2},\partial_{\y_1},\partial_{\y_2}$. To be able to restrict to $M$ we have to make sure that all operators involving $\partial_{\y_1},\partial_{\y_2}$ have coefficients vanishing on $M$.

For example, let us start with the linear combinations
\begin{align}
&\partial_{1} + \partial_{2} = \dfrac{1}{2} \partial_{\x_1} -i \partial_{\y_2}, \\
&\partial_{1} - \partial_{2} =\partial_{\y_1} - \dfrac{i}{2}  \partial_{\x_2}  \, .
\end{align}
We see $\partial_{\y_1},\partial_{\y_2}$ in the r.h.s. so these operators cannot be restricted to $M$. However we can cancel these terms by adding or subtracting antiholomorphic derivatives:
\begin{align}
	&(\partial_{1} + \partial_{2}) + (\bar\partial_{1} + \bar\partial_{2})= \partial_{\x_1}  \, ,\\
	&(\partial_{1} - \partial_{2}) - (\bar\partial_{1} - \bar\partial_{2}) =- i  \partial_{\x_2}  \, .
\end{align}
Note that the terms we add or subtract are just "complex conjugates" of the original operator. We can move the antiholomorphic operators to the r.h.s.:
\begin{align}
	&\partial_{1} + \partial_{2} = \partial_{\x_1} - (\bar\partial_{1} + \bar\partial_{2})  \, \\
	&\partial_{1} - \partial_{2}  =- i  \partial_{\x_2}  + (\bar\partial_{1} - \bar\partial_{2})\, .
\end{align}
Taking the product of these equations we obtain:
\begin{align}
	\partial^2_{1} - \partial^2_{2} = - i  \partial_{\x_1} \partial_{\x_2} +(\bar\partial)  \, ,
\end{align}
where $(\bar\partial)$ is a linear combination of terms each of which contains at least one $\bar\partial_{a}$ and thus vanishes when acting on $\mathcal{G}$. We thus derived~\eqref{eq:op3}.

Analogously consider the following two operators
 \begin{align}
&\k_1 \partial_{1} + \k_2 \partial_{2} = \dfrac{1}{2}\Big(\x_1\partial_{\x_1}+\x_2\partial_{\x_2}\Big)+i\Big(\x_2\partial_{\y_1}-\x_1\partial_{\y_2}\Big) +(M) \, ,
\\ 
&\k_1 \partial_{1} - \k_2 \partial_{2} =  \Big(\x_1\partial_{\y_1}+\x_2\partial_{\y_2}\Big)+\dfrac{i}{2}\Big(\x_2\partial_{\x_1}-\x_1\partial_{\x_2}\Big) +(M) \, ,
\end{align}
where in the r.h.s.~we do not show terms involving differential operators whose coefficients are proportional to $\y_1,\y_2$ and which will vanish when restricted to $M$; such terms are denoted by $(M)$. We now proceed as above, by adding/subtracting the complex conjugate operators, to cancel the terms involving $\partial_{\y_1},\partial_{\y_2}$ in the r.h.s. We obtain:
\begin{align}
	&\k_1 \partial_{1} + \k_2 \partial_{2} = \Big(\x_1\partial_{\x_1}+\x_2\partial_{\x_2}\Big)+(M) +(\bar\partial) \, ,
	\\ 
	&\k_1 \partial_{1} - \k_2 \partial_{2} =  i \Big(\x_2\partial_{\x_1}-\x_1\partial_{\x_2}\Big) +(M)+(\bar\partial) \, .
\end{align}
The first of these equations is~\eqref{eq:op1}, while taking their sum we obtain~\eqref{eq:op2}.

\section{How to Compute the Scattering Rate: Numerical Integration for Ordinary 3D Ising Transition}
\label{app:numint}

This appendix outlines how we computed numerically\footnote{The corresponding {\tt Mathematica} file can be obtained by request from the authors.} the integrals~\eqref{eq:Iy_integral0},~\eqref{eq:Iy_integral}. 
Recall that $|\phi|<\pi/2$. Moreover everything is invariant under $\phi \to -\phi$ so we only need to consider positive $\phi$.

The following basic principles are useful to understand various choices made below: 
\begin{enumerate}
\item
Non-oscillating integrations (like that in $t$) should be done before oscillating ones;
\item
Analytic asymptotics should be combined with numerical integrations in the bulk;
\item
Analytic asymptotics with analytically known integrals may be subtracted, to improve the convergence of numerical integrals.
\end{enumerate}
To follow principle 3, we will determine the asymptotics, see Eq.~\eqref{eq:Ialpha},~\eqref{eq:Ibeta} below\footnote{To avoid misunderstandings, the parameters denoted $\alpha$ and $\beta$ in this appendix have nothing to do with the critical exponents.}
\beq
I(y,\phi) = \begin{cases} C_0(\phi)/y^{\alpha}, &y\to 0\ (\alpha<2)\,,\\
C_\infty/y^{\beta}, &y\to\infty\ (\beta>2)\, ,
\end{cases}
\eeq
and consider
\beq
\Imod(y,\phi) = \frac{C_\infty}{(y^2+a(\phi)^2)^{(\beta-\alpha)/2}y^\alpha}  \, ,
\eeq
with 
\beq
a(\phi)= (C_\infty/C_0(\phi))^{1/(\beta-\alpha)} \, .
\eeq
Then $\Imod$ has the same asymptotics as $I$. Moreover the Fourier transform of $\Imod$ can be computed analytically:
\beq \label{eq:gmod}
\begin{split}
\gmod(p,\phi) = 2\pi\, C_{\infty}
&\left( 
a^{2-\beta }  \frac{\Gamma \left(1-\frac{\alpha }{2}\right) \Gamma \left(\frac{\beta }{2}-1\right) }{2 \, \Gamma \left(\frac{\beta -\alpha }{2}\right)} \,
_1F_2\left(1-\frac{\alpha }{2};1,2-\frac{\beta }{2};\frac{a^2 p^2}{4}\right)+ \right.  \\
&\left.
\left(\frac{p}{2}\right)^{\beta -2} \frac{ \Gamma \left(1-\frac{\beta }{2}\right) }{2 \,\Gamma \left(\frac{\beta }{2}\right)} \, _1F_2\left(\frac{\beta-\alpha}{2};\frac{\beta }{2},\frac{\beta }{2};\frac{a^2 p^2}{4}\right) 
 \right) \,,\quad a=a(\phi)\,.
\end{split}
\eeq
The small $p$ behavior of $\gmod(p,\phi)$ matches the expected small $p$ behavior of $g(p,\phi)$ from App.~\ref{app:smallp}. Indeed we will see that $\beta-2= 2\hD-2$, while the hypergeometric functions are analytic in $p^2$.

Writing the 2-dimensional Fourier transform as a Hankel transform we are thus reduced to evaluating the integral
\beq \label{eq:numint}
2\pi \int_0^\infty dy\, y (I(y,\phi) -\Imod(y,\phi)) J_0(p y) \, ,
\eeq
which will be convergent more rapidly.
At large $y$ we have $(I(y) -\Imod(y)) = O\left(1/y^{\beta+2}\right)$. Using the same arguments as those of App.~\ref{app:smallp} it follows that this integral has a Taylor expansion in $p^2$ up to $O\left((p^2)^{\beta/2}\right)=O(p^{2\hD})$.
This allows us to extract the leading non-analytic behavior of $g(p)$ for small $p$ in closed form, determined by the small $p$ expansion of $\gmod(p)$ and reported in Eq.~\eqref{eq:a11analytic}.

The function $R(u,y,\phi)$ was computed numerically for a discrete set of values of~$\phi$, using its large $u$ asymptotics for $u>15/\cos{\phi}$, and its small $u$ asymptotics at $u=0$.
Using $R(u,y,\phi)$ we computed the $I(y,\phi)$ numerically on a large grid of values of $y$ between $10^{-3}$ and $5000$, using Eq.~\eqref{eq:Iy_integral}. We then computed the integral~\eqref{eq:numint} for a discrete set of momenta, after interpolating $y (I(y,\phi) -\Imod(y,\phi))$ in $y$. Summing the result to $\gmod(p,\phi)$ we obtained the function $g(p,\phi)$, which was used in numerical tests in Section~\ref{sec:num}.

Thanks to the large $y$ behavior $(I(y) -\Imod(y)) = O\left(1/y^{\beta+2}\right)$, we can also extract the small $p$ coefficients 
$e_0(\phi)$ and $e_1(\phi)$ defined in Eq.~\eqref{eq:smallp} by direct integration through
\beq
\begin{split}
&e_0(\phi)=2\pi \int_0^\infty dy y (I(y,\phi) -\Imod(y,\phi)) + e_{0,\rm{model}}(\phi)\, , \\
&e_1(\phi)=2\pi \int_0^\infty dy y (I(y,\phi) -\Imod(y,\phi)) (-y^2/4) + e_{1,\rm{model}}(\phi) \, ,
\end{split}
\eeq
with $e_{k,\rm{model}}(\phi)$ determined by the small $p$ expansion of $\gmod(p,\phi)$, from Eq.~\eqref{eq:gmod}.

\subsection{Asymptotics of $R(u,y,\phi)$ and $I(y,\phi)$}
We derive here the $u$ and $y$ asymptotics of $R(u,y):=R(u,y,\phi)$ and $I(y):=I(y,\phi)$, for a fixed $\phi$. These are used both in the numerical evaluation of the integrals and to determine the  large and small $y$ asymptotics of $I(y)$.

\textbf{Small $u$, fixed $y$.} We can change the integration variable to $\zeta = ut \cos\phi$. From the definition of $R(u, y)$, we find that the limit $u\to 0$ gives 
\beq
 R(0, y) = \dfrac{1}{\cos\phi}\int_0^\infty d\zeta  \,e^{-\zeta} G\left(\dfrac{y^2 \cos^2\phi}{\zeta^2}\right) . 
 \eeq
This tends to $0$ for $y \to \infty$, while it goes to $1/\cos\phi$ for $y \to 0$, interpolating between the two values for intermediate $y$. We use this formula to determine numerically the value of $R(u,y)$ at $u=0$.

\textbf{Small $u$, small $y$.} To find the asymptotic expansion at small $u,y$ it is convenient to subtract the large $t$ behavior of the integrand, which is controlled by the bulk OPE, and treat this analytically:
\beq
\begin{split}
R(u,y)  &=  \,u\int_1^\infty dt \, e^{-t u \cos\phi}   \left(G\left(\frac{1+y^2/u^2}{t^2-1}\right) - 1 - \frac{A(u,y)}{t^{\Depsilon}} \right)  \\
&+ \,u \int_1^\infty dt \, e^{-t u \cos\phi}  \left(1 + \frac{A(u,y)}{t^{\Depsilon}} \right) ,
\end{split}
\eeq
where $A(u,y) = \lambda_1 a_1 (1+y^2/u^2)^{\frac{\Depsilon}{2}}$. 
The first integral admits a Taylor expansion up to order~$(u\cos\phi)^2$, with $y/u$ fixed. We can evaluate its contribution to the term linear in $u \cos\phi$ by setting $u\cos\phi=0$ in the integrand and evaluating this numerically.
For the second integral we can determine its small $u$ expansion analytically. We find
\beq
\dfrac{1}{\cos\phi}\left(e^{-u\cos\phi} + \dfrac{A(u,y)}{\Depsilon-1} u\cos\phi + A(u,y)  \Gamma(1-\Depsilon) (u\cos\phi)^{\Depsilon} + \dots\right) \, .
\eeq

The limit $R(u,y) \to 1/\cos\phi$ for small $u,y$ and fixed $u/y$ is what determines the small $y$ scaling of $I(y)$. Indeed, after changing the integration variable to $\eta= u/y$ in Eq.~\eqref{eq:Iy_integral}:
\beq
I(y)=\frac{1}{y^{2\Ds-1}}  \int_0^\infty  d\eta \frac{\cos(y\eta\sin\phi )}{[1+\eta^2]^{\Ds}} R(y\eta,y).
\eeq
Taking the limit of small $y$ this gives
\beq
I(y) \sim_{y\to0}  C_0(\phi)/y^{\alpha} \, ,
\label{eq:Ialpha}
\eeq
with
\beq
\alpha=2 \Ds-1 \, , \qquad C_{0}(\phi) =\dfrac{1}{ \cos\phi} \times \dfrac{\sqrt{\pi}}{2} \dfrac{\Gamma\left(\Ds-1/2\right)}{\Gamma\left(\Ds\right)} \, .
\eeq

\textbf{Large $u$, arbitrary $y$.} We multiply by $e^{u\cos\phi}$ and change integration variable to ${\zeta= (t-1)u\cos\phi}$:
\beq
e^{u\cos\phi}  R(u,y) = \dfrac{1}{\cos\phi}\int_0^\infty d\zeta \, e^{-\zeta}  G\left(\frac{y^2/u^2+1}{2\zeta/(u\cos\phi)}\right).
\eeq
Expanding for large $u$ but arbitrary $y$ and integrating in $\zeta$ we arrive at:
\beq
R(u,y) \approx  \dfrac{\mu_1^2}{\cos\phi}  \Gamma(1+\hD-\Ds )  \left(\dfrac{2u}{(u^2+y^2)\cos\phi}\right)^{\hD-\Ds} e^{-u \cos\phi}.
\eeq
This approximation is accurate for very large $u$. To get a better approximation we can include the first correction in the large $\xi$ expansion of the boundary OPE channel:
\beq
R(u,y) \approx  \dfrac{\mu_1^2}{\cos\phi}  \Gamma(1+\hD-\Ds )  \left(\dfrac{2u}{(u^2+y^2)\cos\phi}\right)^{\hD-\Ds} \left(1-\dfrac{\hD}{2} \dfrac{2u}{(u^2+y^2)\cos\phi} \right)e^{-u \cos\phi}.
\eeq
This approximation is used in the numerical computation for $u>15/\cos(\phi)$, after including a small correction factor to ensure continuity.

\textbf{Large $y$, fixed $u\ll y$.}
For large $y$ we have:
\begin{align}
R(u,y) & \approx u \int_1^\infty dt  \,e^{-t u \cos\phi} G\left(\frac{y^2}{u^2(t^2-1)}\right)\\
& \approx  \mu_1^2 \left(\dfrac{u}{y}\right)^{2(\hD-\Ds)} u \int_1^\infty dt  \,e^{-t u \cos\phi}  (t^2-1)^{\hD-\Ds} \, ,
\end{align} 
from which 
\beq
R(u,y) \approx_{y \to \infty}  \dfrac{\mu_1^2}{\sqrt{\pi}}  \,\Gamma(1+\hD-\Ds )\, u\left(\dfrac{u}{y}\right)^{2(\hD-\Ds)} \left(\dfrac{2}{u \cos\phi} \right)^{\frac{1}{2}+\hD-\Ds} K_{\frac{1}{2}+\hD-\Ds}(u \cos\phi) .
\eeq
To determine its large $u$ behavior we can use the asymptotics of the Bessel function. We obtain
\beq 
R(u,y) \approx_{\substack{y\to\infty\\ u\to \infty}} \dfrac{\mu_1^2}{\cos{\phi}} \, \Gamma(1+\hD-\Ds)  \left(\dfrac{2u}{y^2 \cos\phi}\right)^{\hD-\Ds}  e^{-u \cos\phi} \, .
\eeq
This agrees with our previous result if $y\gg u$.

Integrating in $u$ the large $y$ expansion of $R(u,y)$, as dictated by Eq.~\eqref{eq:Iy_integral}, we obtain the large $y$ expansion of $I(y)$, after evaluating the integral of the Bessel function. After some nontrivial cancellations we arrive at 
\beq
I(y) \sim_{y\to\infty}  C_\infty/y^{\beta} \, ,
\label{eq:Ibeta}
\eeq
with
\beq
\beta=2 \hD \, , \qquad C_{\infty} = \mu_1^2 \, 4^{\hD-\Ds} \, \Gamma(1+\hD-\Ds )^2 \, .
\eeq
The fact that the leading $y\to \infty$ behavior is independent of $\phi$ is consistent with the fact that the leading non-analytic small $p$ behavior has an angle independent coefficient, $a_{11}$. $a_{11}$ is indeed determined by this asymptotic~\eqref{eq:a11analytic}.

\section{Theoretical terms and abbreviations}\label{sec:glossary}
\begin{description}
	\item[CFT - Conformal Field Theory] A field theory invariant under the action of the conformal group. Various additional properties such as the OPE make it amenable to conformal bootstrap methods, leading to predictions of critical exponents.
	\item[field] (also called ``operator'') A fluctuating $x$-dependent quantity $\O(x)$. The basic observable is a correlation function of a product of several fields at non-coincident points. There are infinitely many fields in any theory, having different scaling dimensions. In a Lagrangian description, some of these will be fundamental and some composite.
	\item[OPE - Operator Product Expansion] Expansions of field theory correlation functions, obtained by replacing (the product of) a group of fields in a correlator by an infinite series of other fields. In this paper boundary and bulk OPE expansions were used.
    \item[operator] See ``field''.
	\item[primary] Fields which are not total derivatives (of some order) of other fields. 
	\item[RG - Renormalization group] The study of renormalization group transformations and their fixed points. Historically the first general method to compute critical exponents. It does not rely on conformal invariance.
	\item[scaling dimension] A real number $\Delta>0$ characterizing the transformation of a field under scale transformations. It fixes the power in the dependence of a two-point correlation function from the distance. In a conformally invariant theory, the same number also fixes the conformal transformation properties of the field.
	\item[two-point function] Correlation function of a product of two fields.
	\item[WI - Ward identity] A generic term for any differential equation expressing invariance of a correlation function under an infinitesimal symmetry transformation. Can be formulated in position or in momentum space. Originally derived in the context of Quantum Electrodynamics (QED), they are also known as Ward-Takahashi identities.
\end{description}

\bibliography{biblio.bib}
\bibliographystyle{utphys}

\end{document}